\documentclass[aps,pre,preprint,groupedaddress,showkeys]{revtex4-1}

\usepackage{amssymb}
\usepackage{graphicx}
\usepackage{color}

\begin{document}

\preprint{http://arxiv.org/abs/1506.08129}

\title{Fluorescence correlation spectroscopy in thin films at reflecting substrates as a means to study nanoscale structure and dynamics at soft-matter interfaces}

\author{Daniela T\"auber}
\email{dantaube@gmx.de}
\homepage{www.daniela-taeuber.de}
\altaffiliation{now at Department of Chemical Physics, Lund University, S-22362 Lund, Sweden}
\author{Kathrin Radscheit}
\affiliation{Institute of Physics, Technische Universit\"at Chemnitz, D-09107 Chemnitz, Germany}
\author{Christian {von Borczyskowski}}
\affiliation{Institute of Physics, Technische Universit\"at Chemnitz, D-09107 Chemnitz, Germany}
\author{Vladimir Al.\ Osipov}
\affiliation{Department of Chemical Physics, Lund University, S-22362 Lund, Sweden}
\author{Michael Schulz}

\affiliation{Institute of Physics, Technische Universit\"at Chemnitz, D-09107 Chemnitz, Germany}


\date{\today}

\begin{abstract}
Structure and dynamics at soft-matter interfaces play an important role in nature and technical applications. Optical single-molecule investigations are non-invasive and capable to reveal heterogeneities at the nanoscale. In this work we develop an autocorrelation function (ACF) approach to retrieve tracer diffusion parameters obtained from fluorescence correlation spectroscopy (FCS) experiments in thin liquid films at reflecting substrates. This approach then is used to investigate structure and dynamics in 100~nm thick 8CB liquid crystal films on silicon wafers with five different oxide thicknesses. We find a different extension of the structural reorientation of 8CB at the solid-liquid interface for thin and for thick oxide. For the thin oxides, the perylenediimide tracer diffusion dynamics in general agrees with the hydrodynamic modeling using no-slip boundary conditions with only a small deviation close to the substrate, while a considerably stronger decrease of the interfacial tracer diffusion is found for the thick oxides.
\end{abstract}

\pacs{68, 78, 81, 82}

\maketitle

\section{Introduction}
Understanding nanoscale dynamics at soft-matter interfaces is of profound interest in life sciences as well as for technical applications. For example, nanoscale structure and dynamics are crucial for the performance of organic solar cells~\cite{vakhshouri_characterization_2013, deibel_organic_2010, camacho_polarization_2014} and inkjet printed electronics~\cite{belgardt_inkjet_2013}, while, for example, protein adsorption is of great interest in life sciences~\cite{schmitt_structural_2010}, and in particular plays an important role in the biocompatibility of implants~\cite{thevenot_surface_2008}. The application related research is accompanied by fundamental research on the influence of structural changes at soft-matter interfaces on material properties\cite{napolitano_glassy_2013, mcumber_electrostatic_2015}.
Nowadays several well established investigation methods with nanoscale and even atomic resolution exist, for instance, x-ray reflectometry~\cite{tolan_x-ray_1999} and atomic force microscopy~\cite{mcconney_probing_2010, spitzner_multi-set_2012}. However, their application to the investigation of dynamics in soft-matter is still delicate, as they usually demand for sophisticated sample preparation and may cause damage to the materials of interest. The fortune of optical investigation methods is their non-invasiveness. Thereby, single molecule methods are superior to ensemble methods in retrieving heterogeneous structures and dynamics on the micro- and nanoscale~\cite{kulzer_single_2010, woll_polymers_2009}. Fast diffusion processes and other dynamics related to fluorescence fluctuations in the range from nanoseconds to seconds can be monitored by fluorescence correlation spectroscopy (FCS)~\cite{elson_fluorescence_1974, schwille_fluorescence_2002, woll_polymers_2009, tauber_influence_2013}. Conventional FCS is retrieved employing a confocal fluorescence microscope~\cite{elson_fluorescence_1974, schwille_fluorescence_2002}. Several more advanced microscope techniques have been developed during the recent years~\cite{sauer_handbook_2010, engelborghs_fluorescence_2014}, including two-photon excitation for FCS and also a combination of FCS with stimulated emission depletion (STED)~\cite{vicidomini_sharper_2011}. In contrast to many other super resolution techniques, STED does not require for statistical averaging, enabling also a high temporal resolution for investigation of fast dynamics~\cite{leung_review_2011}. For example, it has been successfully applied to discriminate between free and hindered diffusion of lipids in a living cell, which had not been possible by conventional confocal microscopy~\cite{eggeling_direct_2009}. However, drawbacks of STED are its need for alignment of the excitation and depletion beams, and a dependence of the spatial resolution on relatively high excitation powers~\cite{leung_review_2011}, which requires very photostable fluorophores and may damage the sample structure. 

Conventional confocal microscopy provides a vertical resolution in the range of 0,5 to 1~$\rm\mu m$~\cite{resch-genger_state_2008, hein_stimulated_2008}, it may be further reduced by using more sophisticated methods~\cite{engelborghs_fluorescence_2014}. For example, 3D STED allows to achieve a vertical resolution of 150~nm~\cite{hein_stimulated_2008}. Here we demonstrate a reduction of the vertical resolution in conventional confocal microscopy down to 50-100~nm by employing optical interference at mirror interfaces. The approach does not require any alignment of two or even more laser beams, and it is particularly suited to retrieve information about the structure and dynamics at soft-matter interfaces. In principle, more advanced methods may also be employed together with mirror interfaces, however, the implications of the wavelength dependent interference patterns on such methods have to be taken into account. Rigneault and Lenne developed modifications of 3D correlation functions for FCS at mirror interfaces~\cite{lenne_subwavelength_2002, rigneault_fluorescence_2003, etienne_confined_2006}. Below we address the problem of diffusion in films which are considerably thinner than the focal depth of the excitation profile. Investigation of  thin films on this scale is of particular interest for understanding the structure and physical properties at soft-matter interfaces. We apply the developed correlation function to FCS studies of 100~nm thick liquid crystal (LC) films on silicon substrates with varied silicon oxide thicknesses $d$. 

Liquid crystal (LC) materials have been of increasing interest within the recent decades, on the one hand due to their abundance in living organisms~\cite{cowin_liquid_2004, giraud-guille_liquid_2008}, and on the other hand due to an increasing field of technical applications, which does not only span liquid crystal displays and optoelectronic devices. For example, recently, LC materials have been investigated for enhancement of organic solar cell performance~\cite{suzuki_effect_2014}. The here employed 4-n-octyl-4'-cyanobiphenyl (8CB) is in its smectic-A state at room temperature and allows us to compare the results with literature reports on investigations using different experimental methods~\cite{tauber_single_2013, coursault_self-organized_2015}. LC materials are very sensitive not only to interface structures, but also to interaction forces with the substrate, influencing their anchoring conditions~\cite{de_gennes_physics_2002, lin_note_2013}. For the here studied 8CB films, in particular, an influence of the $\rm SiO_2$ layer thickness $d$ on film structure and dynamics was found~\cite{schulz_influence_2011, tauber_guest_2013}. Here, we vary $d$ in five steps from native (4~nm) to 100~nm, which we expect to shed some light onto this influence. 

To allow for a general overview of this study, we start with a short outline of the experiment, before going into details about the developed fluorescence correlation function in section~\ref{FCSfunction} and the calculation of the vertical fluorescence modulation in section~\ref{flumod}.
\vspace{1in}

\section{\label{Scope}Scope of the experiment}
\begin{figure}[ht]
 \begin{minipage}[b]{3.4in}
 \includegraphics[width=2.8in]{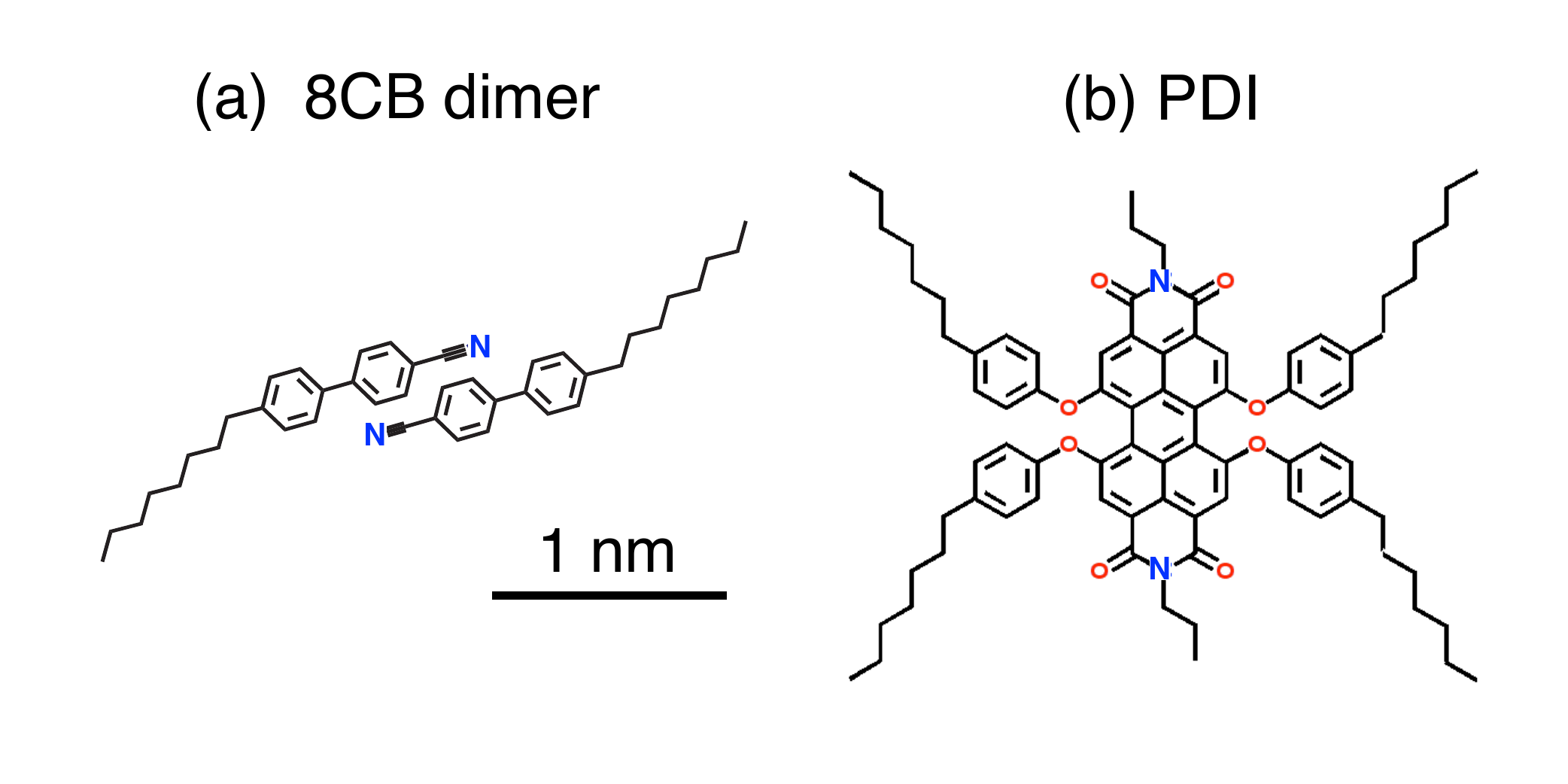}
  \includegraphics[width=1.5in]{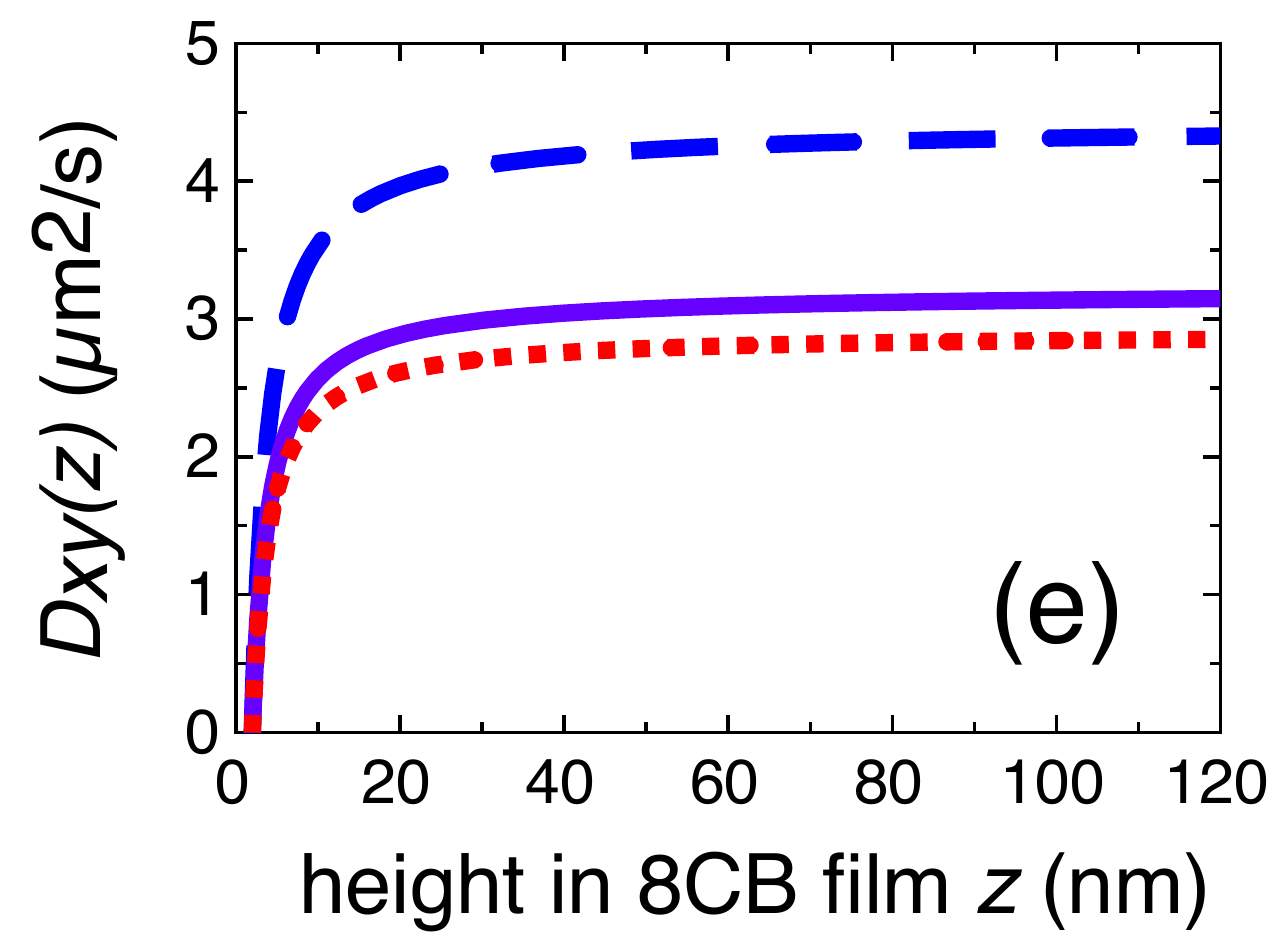}
  \includegraphics[width=1.5in]{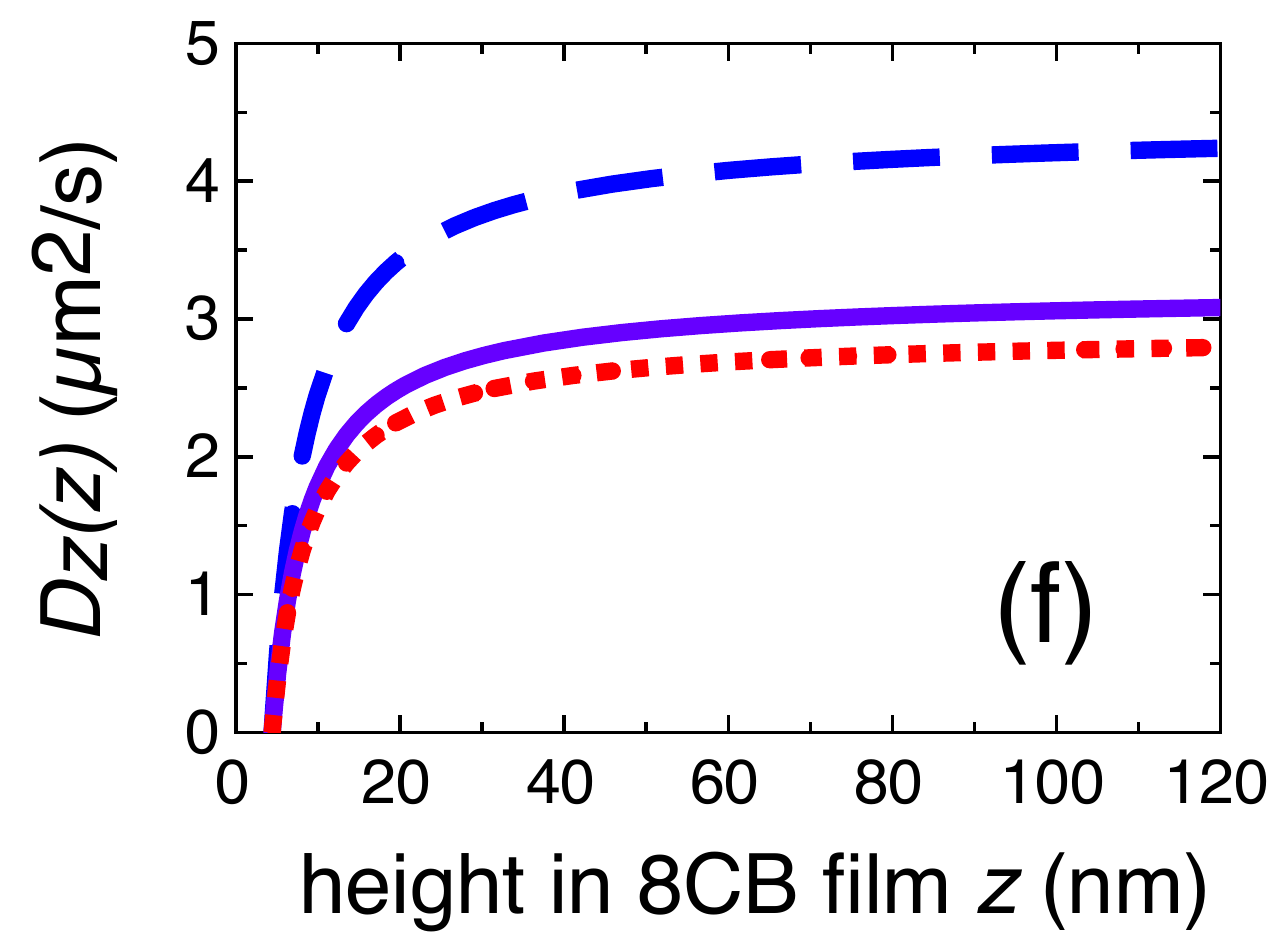}
 \end{minipage}
 \hspace{0.2in}
 \includegraphics[width=1.1in]{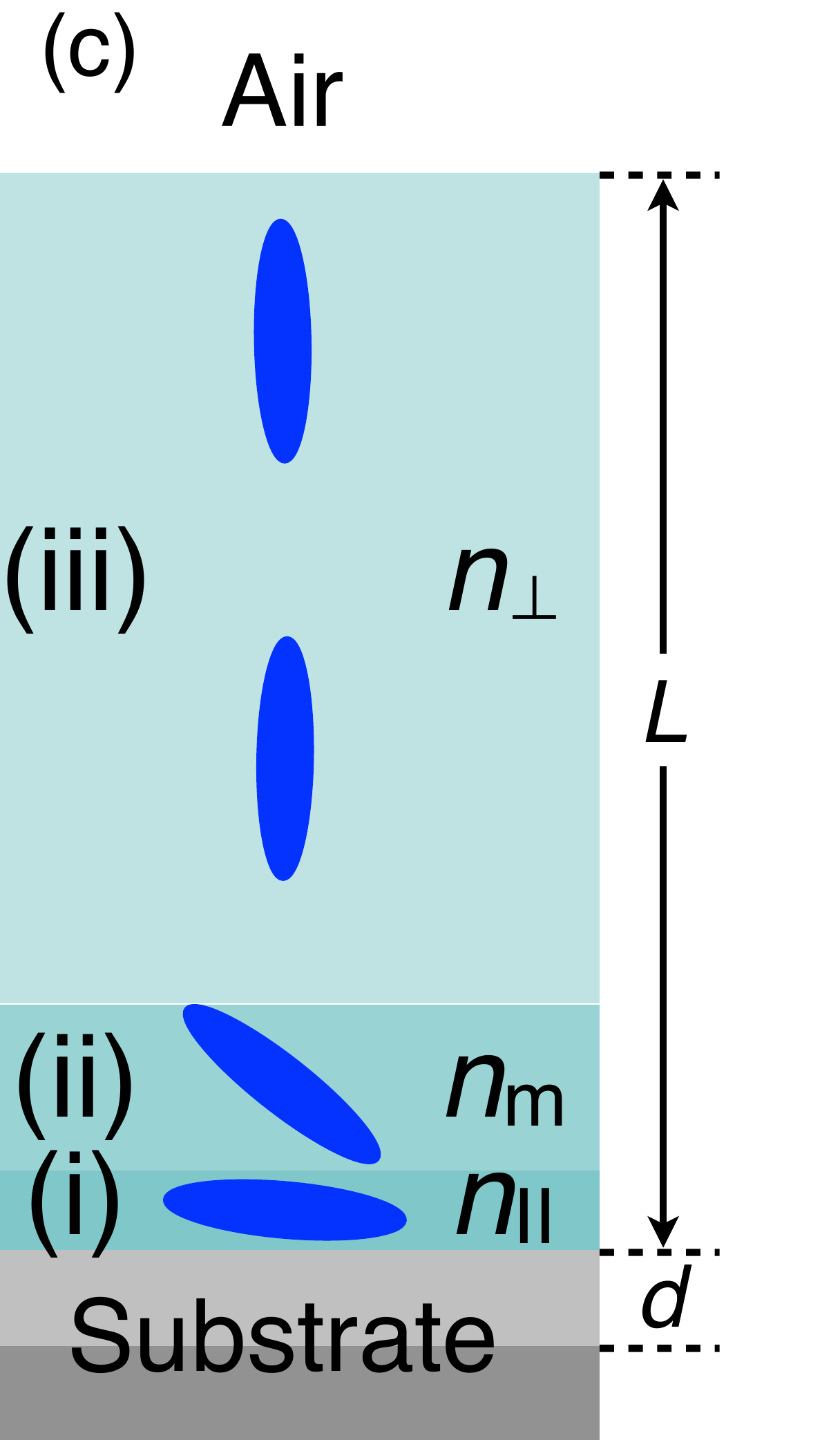}
 \includegraphics[width=1.1in]{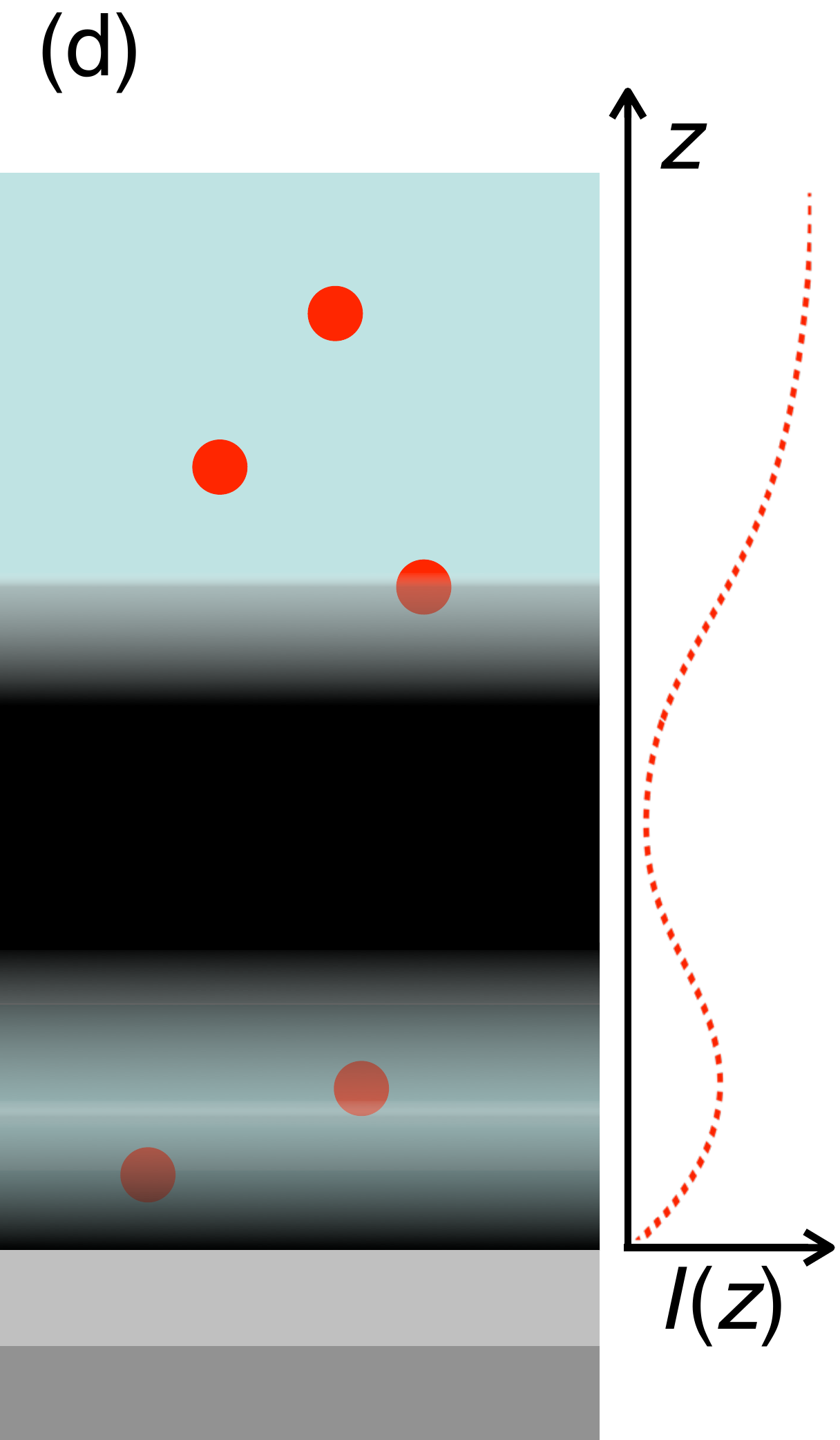}
\caption{\label{structures} a,b) Chemical Structures. c) Model (not to scale) of sample structure with 8CB film thickness $L$, $\rm SiO_2$ layer thickness $d$; with refractive indices changing from $n_{\parallel}$ at the substrate (region i) to $n_{\perp}$ at the interface with air (region iii) and an intermediate region (ii) with mean refractive index $n_{\rm m}$. Orientation of 8CB dimers highlighted as blue cigars. d) Scheme illustrating the use of the interference pattern in $I(z)$ for probing the samples, no fluorescence signal from PDI tracers (red balls) in dark regions. e,f) Diffusion coefficients according to hydrodynamics: e) $D_{xy}(z)$ and f)  $D_{z}(z)$, calculated from (\ref{Dpll_noslip}) and (\ref{Dperp_noslip}), respectively, for mean diffusion (solid), as well as diffusion along (dash) and perpendicular (dot) to the LC director. }
\end{figure}
In this work we study thin 8CB films of thickness $L=110\pm10$~nm on silicon substrates differing in oxide thickness $d$. For details on sample preparation and conduction of measurements we refer to appendix \ref{Experiment}. As stated above, the particular structure of the 8CB film is very sensitive to interface structure and forces~\cite{de_gennes_physics_2002, lin_note_2013, schulz_influence_2011, tauber_guest_2013}. In bulk 8CB the molecules will arrange paired into dimers with antiparallel dipole moments~\cite{xu_wetting_2000}, see FIG.~\ref{structures} a) and the LC director oriented parallel to the long axis of the 8CB dimers. We modeled the vertical structure within the 8CB films using three regions differing by the orientation of these 8CB dimers, as shown in FIG.~\ref{structures} c), with the 8CB dimers highlighted as blue cigars (not to scale). As 8CB is birefrigent, the refractive index depends on the particular orientation and is noted also in FIG.~\ref{structures} c), ranging from parallel to the LC director $n_{\parallel}$ close to the substrate, to perpendicular to the LC director $n_{\perp}$ at the interface with air, with an intermediate region for which we employ the mean reflective index $n_{\rm m}$. Further details on the film structure are discussed in section~\ref{flumod} on the vertical fluorescence modulation.

For investigating structure and dynamics in these thin 8CB films, we used highly diluted perylenediimide (PDI) molecules as fluorescence tracers, for the chemical structure see FIG.~\ref{structures}b). The PDI tracers are depicted as red balls in FIG.~\ref{structures}d), which illustrates the fluorescence detection depending on the vertical fluorescence intensity modulation $I(z)$. The dark regions correspond to low values of a typical $I(z)$ caused by negative interference. Fluorescence from diffusing PDI tracers can only be detected in the bright regions, which depend on $d$ and differ for the two employed excitation wavelengths $\lambda_{ex}=488$~nm and $\lambda_{ex}=515$~nm. Numerical data for each calculated $I(z)$ were employed with the below described fluorescence intensity autocorrelation function (ACF) to derive vertical $D_z$ and lateral $D_{xy}$ diffusion coefficients. Thereby, the vertical dependence of the diffusion coefficients was modeled by a potential of strength $V_0$ rapidly decaying with distance $z$ from the substrate, as will be discussed in more detail in the next section~\ref{FCSfunction} and in appendix~\ref{DerivCorr} giving details on the derivation of the ACF. 

Since the detection of diffusing tracer molecules is restricted to the areas with sufficiently high $I(z)$ (bright areas in FIG.~\ref{structures}d), the experimental FCS data report only on the corresponding vertical regions of the film. By incorporating $I(z)$ into the particular ACF for fitting the data, we take this into account. Consequently, the obtained potential and diffusion coefficients are related to the particular vertical region in the films spanned by the combination of all $z$ for which $I(z)\gtrsim 0.4$. As this not necessarily covers the whole film thickness $L$, we employ a hydrodynamic model with no-slip boundary conditions for calculating height dependent diffusion coefficients and comparing them with our results. Details about this hydrodynamic model are given in appendix~\ref{hydro}. The calculated vertical dependence of the diffusion coefficients is plotted in FIG.~\ref{structures}e) and f) for $D_{xy}$ and $D_z$, respectively. Thereby, the diffusion anisotropy of 8CB is taken into account by plotting diffusion coefficients along (dash) and perpendicular (dot) to the LC director, as well as mean diffusion coefficients (solid).

Before discussing the results from our investigation, we will explain the employed ACF and its derivation in the following section~\ref{FCSfunction} as well as the calculation of $I(z)$ in section~\ref{flumod}.

\section{\label{FCSfunction}Fluorescence intensity correlation function}
In general, for FCS, a temporal (auto-)correlation of the fluorescence intensity ($\mathcal{F}$) or the related (auto-)correlation of the fluorescence intensity fluctuations ($\delta \mathcal{F}=\mathcal{F}-<\mathcal{F}>$) recorded from a confocal volume in the sample is required~\cite{elson_fluorescence_1974, schwille_fluorescence_2002}. As long as no other fluorescence modulation mechanisms in the time region of interest are involved, the diffusion coefficients of the fluorescing probes diluted in the sample may be derived from the ACF. 

The ACF for translational diffusion in FCS experiments is based on the analysis of the expression~\cite{schwille_fluorescence_2002}
\begin{equation}
<g(t,t^{\prime })>\equiv<\delta \mathcal{F}(t)\delta \mathcal{F}(t^{\prime })>=\int \int \limits_{G}d^{3}r\ d^{3}r^{\prime }\ I(\mathbf{r}%
)\left\langle \varrho (\mathbf{r},t)\varrho (\mathbf{r}^{\prime },t^{\prime
})\right\rangle I(\mathbf{r}^{\prime })\, .  
\label{start}
\end{equation}%
Here, $I(\mathbf{r})$ is the normalized spatial distribution of the detectable light intensity fluctuations ($\delta \mathcal{F}=I(\mathbf{r})\varrho (\mathbf{r},t)$), the function $\left\langle
\varrho (\mathbf{r},t)\varrho (\mathbf{r}^{\prime },t)\right\rangle $ is the
density-density correlation function of the active particles in the detection volume $G$, and the average is taken over the full time of observation. For derivation of the ACF, on the one hand, the particular $I(\mathbf{r})$ in the sample has to be known, and on the other hand, knowledge about the expected diffusion process is needed. Typically, FCS is used in bulk biological samples, where the spatial behavior of $I(\mathbf{r})$ can be approximated by a 3D Gaussian~\cite{schwille_fluorescence_2002}, and the density of the active particles is expected to be homogeneous. Then the normalized ACF depends on the ratio of the lateral Gaussian waist $w_{xy}$ to the vertical one $w_z$ and has the form~\cite{schwille_fluorescence_2002}
\begin{equation}
g_{3D}(\tau) = \frac{1}{(1+\tau/\tau_D)\sqrt{1+(w_{xy}/w_z
)^2(\tau/\tau_D)}}\, .
\label{3Dcorr}
\end{equation} 
where the correlation time $\tau$ is scaled with the correlation time for translational diffusion $\tau_D$. If FCS is applied to films considerably thinner than the focal depth $w_z$, the vertical diffusion of the fluorescent probes does not contribute to the ACF. In this case one uses a 2D Gaussian approximation for the lateral shape of the fluorescence intensity distribution $I(\mathbf{r})$~\cite{elson_fluorescence_1974, cooper_imaging_2014}, which eventually leads to the 2D ACF
\begin{equation}
g_{2D}(\tau) = \frac{1}{1+\tau/\tau_D}\, .
\label{2Dcorr}
\end{equation} 
In both cases, the ACF~(\ref{3Dcorr}), (\ref{2Dcorr}) will yield diffusion coefficients averaged over the focal region within the sample during the observation time. 

By use of a mirror interface, the sampling region can be further reduced due to the interference pattern, which for visual light typically yields interference fringes on a scale of $50-100$~nm~\cite{rigneault_fluorescence_2003, etienne_confined_2006, lenne_subwavelength_2002}. Assuming a cosine fringe pattern within the focal depth, a simple extension of the 3D ACF~(\ref{3Dcorr}) may be used to obtain the correlation function in the form~\cite{lenne_subwavelength_2002}
\begin{equation}
g_f(\tau) = g_{3D}(\tau)\left[1+A\,{\rm exp}\left(- \frac{\tau}{\tau_f}\right)\right]\, ,\quad\mbox{with}\quad 
\tau_f=\frac{\lambda_{ex}^2}{16\pi^2n^2D}\,,
\label{fringe}
\end{equation} 
where $A$ is a constant, $\lambda_{ex}$ is the excitation wavelength, $n$ is the refractive index of the solvent, and $D$ is the translational diffusion coefficient. Rigneault and Lenne successfully applied eq.~(\ref{fringe}) to analyze diffusion of Cy5 in an 80~$\rm\mu m$ thick water layer on a dielectric mirror interface~\cite{rigneault_fluorescence_2003}. They further extended this approach for diffusion in small tubes~\cite{etienne_confined_2006}. 

Eq.~(\ref{fringe}) can be used as long as homogeneous diffusion within the sample volume is assumed. Structural changes within the sample, in particular close to the interface, as well as interactions of the fluorescent probe with the substrate cause heterogeneities~\cite{tauber_influence_2013, tauber_characterization_2011}. If those heterogeneities cannot be neglected, a more precise calculation of the vertical excitation and fluorescence detection is required, together with an advanced model of the diffusion of the fluorescent probes. On the other hand, this very interplay between sample structure, resulting vertical fluorescence modulation and diffusion coefficients can be utilized to retrieve structure related diffusion coefficients, and thus to increase the amount of information  obtained from vertical regions in the sample on a scale of 50 to 100~nm.

Here we propose an ACF approach developed for the diffusion of fluorescent probes in thin films on mirror interfaces, when the film thicknesses $L$ is considerably smaller than the focal depth $w_z$. Contrary to the previously discussed cases, the vertical  fluorescence intensity distribution $I(z)$ (the zero of $z$-axis, is placed at the interface of the substrate with the LC film, the positive direction is chosen into the thickness $L$ of the LC film) is no longer Gaussian, but is modified by the optical interference (see section~\ref{flumod}), causing a vertical modulation of the Gaussians in the $x,y$-plane, $I(\mathbf{r})=I_0 I(z)\exp\left\{-\frac{x^2+y^2}{2w_{xy}^2}\right\} $ (the zero point $x=0$, $y=0$ is placed at the geometrical center of the laser beam). 

The diffusion takes place in an anisotropic LC which generates a diffusion anisotropy, where diffusion parallel to the LC director ($D_{\parallel}$) is $\approx 1.5$ times faster than diffusion perpendicular to it ($D_\perp$)~\cite{tauber_single_2013}. The geometry of the experimental focal volume (see appendix \ref{Experiment}) with $2w_{xy}\approx 0.5$~$\mu{\rm m}>L\approx 0.1$~$\mu{\rm m}$ yields vertical correlation times at least 20 times shorter than lateral ones. This allows to distinguish the diffusion processes taking place in vertical and lateral directions. Due to the diffusion anisotropy of the material it is feasible to employ this experimental possibility and to discern both corresponding diffusion coefficients, the vertical $D_z$ and the lateral diffusion coefficient $D_{xy}$. The LC director orientation at the boundaries is known from literature~\cite{designolle_afm_2006}, and one expects that $D_z(z)=D_{\parallel}$ and $D_{xy}(z)=D_{\perp}$ as  $z\to L$, while this does not hold for $z\to 0$. The interaction of the LC with the substrate leads to a $z$-dependence of both $D_z(z)$ and $D_{xy}(z)$, which is modeled by a $z$-dependent drift term in the diffusion equation, generated by a potential $V(z)$ which rapidly decays with the distance from the substrate (as $z$ increases), for details about the mathematical model see appendix~\ref{DerivCorr}.

It is clear that the choice of the potential $V(z)$ significantly depends on the geometry of the system. According to Blake, diffusion of a liquid close to a wall with hydrodynamic no-slip boundary condition can be represented by a potential with a square decay with the distance to the wall~\cite{blake_note_1971}. For the here studied thin LC films, a hydrodynamic no-slip boundary condition at the substrate is feasible~\cite{schulz_optical_2010}. Additionally, the LC properties together with the confined geometry give rise to so called pseudo casimir forces~\cite{ajdari_pseudo-casimir_1992}, which can also be represented by a $1/z^2$ decaying potential. Besides the influence of the confinement on the self-diffusion of the LC molecules, the tracer molecules themselves will interact with the substrate.  For the here used neutral perilenediimide tracer molecules, hydrogen bonding with substrate surface silanols is the most likely tracer-substrate interaction~\cite{tauber_characterization_2011}. According to Israelachvili, the potential of hydrogen bonds between two molecules decays with the square of the distance and contains also a dependence on the bond angle~\cite{israelachvili_intermolecular_2011}. In our system the interface has an unknown irregular distribution of silanols and thus unknown bond angles, in addition, the interaction potential might be screened by the LC medium, leading to an even faster decay. Long range van der Waals forces acting between the tracer molecules and the substrate decay with the cube of the distance~\cite{parsegian_van_2006}.

As the result of this discussion we model the potential as 
\begin{equation}
V(z)=V_0/z^2\, 
\label{potential}
\end{equation}
with some potential strength $V_0$. Rapid decrease of the potential allows to employ perturbation theory for solving the diffusion equation, for details see appendix~\ref{DerivCorr}. Thereby, the diffusion coefficients are approximated as being homogeneous within the area of observation, while their $z$-dependence is taken into account by the potential. The other simplifying assumptions are that the diffusing isotropic perylenediimide tracers are highly diluted and that one can use the reflecting boundary conditions at both interfaces. These assumptions are supported by previous single molecule tracking experiments, which in particular showed only a small amount of adsorption events ($<17$\%)~\cite{schulz_optical_2010}. 
 
Besides the above explained impacts from the confinement on tracer diffusion at solid-liquid interfaces, an influence from the particular LC structure is also expected. In LC materials, the interfacial interactions cause distortions of the bulk LC structure which itself is induced by the intermolecular interactions. The particular anchoring conditions at the substrate depend on the substrate chemistry and roughness~\cite{mullin_properties_1989, pizzirusso_predicting_2012, roscioni_predicting_2013}, while the anchoring strength also depends on the long range interactions~\cite{schulz_influence_2011,  sarlah_van_2001, ajdari_pseudo-casimir_1992}. For thin nematic LC films with antagonistic anchoring, i.e.~perpendicular (homeotropic) anchoring at the air-interface and planar anchoring at the substrate, Lin et al.~showed that strong anchoring stabilizes the film structure~\cite{lin_note_2013}, while weak anchoring leads to instabilities of the film thickness and the appearance of LC director distortions~\cite{effenterre_coupling_2003, lin_note_2013}. For the here studied thin smectic-A LC films with antagonistic anchoring conditions, the extent of reorientation at the substrates is still not fully understood and thus is not included into the modeled potential~(\ref{potential}). The expected influence from the LC reorientation will be discussed along the presentation of our results.

\section{Vertical fluorescence modulation \label{flumod}}
To apply the above derived correlation function to the investigation of thin films at reflecting substrates, the particular vertical fluorescence intensity modulation has to be calculated. Naturally, it is related to the refractive indices of the involved media. Since the here employed 8CB is birefringent, its arrangement within the thin film has to be modeled, as we briefly stated in section~\ref{Scope} and depicted in FIG.~\ref{structures} c). We now will discuss details of the film structure and the employed model before explaining the calculation of the vertical fluorescence modulation.

In the smectic~A phase the 8CB dimers form layers spaced by about 3.2~nm aligning perpendicular to the layers~\cite{xu_wetting_2000}. At silica substrates 8CB molecules are known to align tilted planar, with a tilt angle of $\approx23^\circ$~to the interface plane~\cite{mullin_properties_1989}, while they will align perpendicular (homeotropic anchoring) at the interface with air~\cite{designolle_afm_2006}. For smectic liquid crystals free energy considerations are strongly unfavorable in respect to bending of the smectic layers~\cite{de_gennes_physics_2002, lacaze_bistable_2004}. Thus, the arrangement of 8CB in thin films with such antagonistic anchoring conditions is not straightforward~\cite{designolle_afm_2006}. Just recently, investigations of thin 8CB films on rubbed PVA were reported~\cite{coursault_self-organized_2015}. On rubbed PVA, 8CB aligns unidirectional planar, in contrast to the randomly oriented planar anchoring at amorphous silica. But also there the mismatch of anchoring conditions on the substrate and on the air-interface, causes deformations of the smectic LC film. The authors suggest a transition zone to form close to the substrate that ranges between 27 and 49~nm into the film, while it may contain a number of dislocations or a melted nematic area~\cite{coursault_self-organized_2015}. On silicon with native oxide, for very thin 8CB films a trilayer conformation is reported~\cite{xu_wetting_2000}. For thicker films, the formation of a nematic layer at the interface with the substrate is suggested~\cite{lacaze_bistable_2004, schulz_influence_2011}.
Recently Roscioni et al.\ performed atomistic molecular dynamics simulations of the 8CB homolog 4-n-pentyl-4'-cyanobiphenyl (5CB) at crystal as well as amorphous silica interfaces~\cite{roscioni_predicting_2013}. For nematic 5CB on the amorphous silica reorientation from random planar to perpendicular orientation occurred within 5 to 15 nm from the substrate in 20 to 25~nm thick films~\cite{roscioni_predicting_2013}. For our considerably thicker smectic LC films with $L=110$~nm, we expect the reorientation to extend further into the film. 
Preliminary studies of 220~nm thick 8CB films on silicon with native or 100~nm thick oxide showed an influence of the oxide thickness on the film structure, suggesting stronger anchoring for 8CB on native oxide~\cite{schulz_influence_2011, tauber_guest_2013} and thus a varying extent of the reorientations depending on the oxide thickness $d$. 
For methodical reasons, we choose to use three different extensions of the interfacial order spaced by 10~nm, namely 15, 25 and 35 nm from the substrate into the film to calculate possible fluorescence intensities $I(z)$ and compare the related fit results.
To approximate the random planar arrangement at the substrate and the following reorientation, we split the modeled interfacial region into two parts: (i) close to the substrate a 3-5~nm thick region with molecules aligned planar and refractive index $n_\parallel$, (ii) followed by a region of rearrangement for which we use the mean refractive index $n_{\rm m}$. The remaining top region (iii) of the film consists of smectic layers parallel to the substrate with refractive index $n_\perp$, see FIG~\ref{structures}~c). In the following we refer to the combination of regions (i) and (ii) as disordered region with height $\zeta$, motivated by the circumstance that at $22^\circ$~C bulk 8CB is expected to be in its smectic~A phase, in contrast to this interfacial rearrangement. 

To obtain $I(z)$ we used {\it Essential Macleod}\,\texttrademark \, (V.9.7 Thin Film Center, Tucson Arizona) to calculate 
the intensity profiles $I_j(\varphi ,\xi)$ within our thin films for relevant wavelengths $j$, whereby $\xi$ is the vertical 
position in respect to the reflecting silicon interface and $\varphi$ is the incident and reflected angle at the film surface. 
{\it Essential Macleod}\,\texttrademark\, allows to use uniaxially birefringent media within the calculation of reflections in thin films. 
The fluorescent probe PDI does not align with the LC, but is isotropically distributed~\cite{schulz_optical_2010}.
 Within 8CB its fluorescence spectrum contains a peak at $\lambda_{peak}=620$~nm and a $60\%$ intense shoulder peak 
at $\lambda_{shoulder}=647$~nm~\cite{schulz_optical_2010}. 
We model the intensity in emission $I_{em}(\varphi ,\xi)$ as a linear combination of the intensity 
profiles $I_{\rm peak}(\varphi ,\xi)$, $I_{\rm shoulder}(\varphi ,\xi)$ for $\lambda_{peak}$, $\lambda_{shoulder}$, respectively, 
and use it together with the intensity profiles in excitation $I_{ex}(\varphi ,\xi)$ to obtain angular fluorescence profiles 
$I(\varphi ,\xi)$ for the two excitation wavelengths, 465~nm and 514~nm according to 
\begin{equation}
I(\varphi ,\xi) = I_{ex}(\varphi ,\xi) \left[I_{\rm peak}(\varphi ,\xi)+0.6\,I_{\rm shoulder}(\varphi , \xi)\right]\, .
\label{fluoprofile}
\end{equation} 
The numerical aperture of 0.9 corresponds to a maximum angle of $64^\circ$. Thus, $I(z)$ was obtained by setting $z=\xi-d$, 
where $d$ is the oxide layer thickness of the substrate and averaging the intensity $I(\varphi ,\xi)$ over the area 
illuminated by the objective using the approximation 
\begin{equation}
I(z) = \frac{1}{{\sin^2(64.5^\circ)}}\left(\sin^2(0.5^\circ)I(0^\circ,z)+ \sum_{\varphi=1}^{64}\left[\sin^2(\varphi+0.5^\circ)-\sin^2(\varphi-0.5^\circ)\right]I(\varphi,z)\right)
\, .
\label{filmarea}
\end{equation} 
The above expression is a discrete integral (discretization step $\varphi=1^\circ$) 
of $I(\varphi,z)$ over the surface of the LC air interface in polar coordinates. The fluorescence intensity from a tracer molecule reaching the LC air interface will be decreased by an increasing path length of incident and emitted light in the LC medium upon increasing the angle $\varphi$~\cite{lakowicz_principles_2010}. We approximated this drop of $I(\varphi,z)$ to be proportional to $\cos^2\varphi$.
Essential Macleod\texttrademark\, provides $I(\varphi ,\xi)$ for $s$, $p$ and a mean polarization. Here 
we used the mean polarization, because our excitation was circularly polarized, and the isotropic dye 
molecules could freely rotate~\cite{schulz_optical_2010}. The obtained vertical fluorescence intensity modulations $I(z)$ 
do only contain averages over illumination and detection angels. The lateral gaussian intensity distribution was already 
taken into account for the derivation of the correlation function (\ref{zerog},~\ref{firstg}) and thus is not included here. 

In close proximity to silicon ($\xi<20$~nm), $I(z)$ is not only determined by interference, but also by non-radiative fluorescence quenching~\cite{hayashi_quenching_1983}. Several mechanisms for this quenching have been discussed in literature. According to Danos et al., an exponential distance dependence of the ratio of non-radiative to radiative de-excitation will fit most experimental data well~\cite{danos_efficient_2008}. For $d\leq25$~nm we further applied this approximation to $I(z)$ from eq.(\ref{filmarea}) to obtain the final $I(z)$.

\section{Results and Discussion}
As outlined above, diffusion dynamics in liquids close to interfaces with solids deviate from the bulk behavior. Moreover, for the here studied 8CB films, the diffusion dynamics are related to the - yet not fully understood - local LC structure, and we expect to see an influence from the local structure in our results. In this study we compare FCS experiments from 10 different experimental conditions, where the optical interference leads to detection of the FCS signal from  specific vertical regions in the films, which are varied by changing between two excitation wavelengths $\lambda_{ex}$ and five oxide layer thicknesses~$d$. This section is therefore structured as follows: In part~\ref{calculated_I} we demonstrate the influence from the reflecting substrates on the FCS experiment by comparing the obtained curves from such experiments to those on nonreflecting glass substrates, and we give a comparative overview of the calculated vertical fluorescence intensity modulations $I(z)$. For a detailed discussion of the  experimental FCS curves for the various combinations of $\lambda_{ex}$ and $d$ as well as the fit results from application of the derived correlation function~(\ref{zerog},~\ref{firstg}) we refer to appendix~\ref{FitsExp}. In two further parts we summarize and discuss the obtained results for the LC structure in the thin 8CB films (part~\ref{LC_disorder}), and for the diffusion dynamics (part~\ref{diff_dyn}).

\subsection{Experimental FCS curves and calculated vertical fluorescence modulations\label{calculated_I}}
\begin{figure}[htb]
 \includegraphics[width=2.49in]{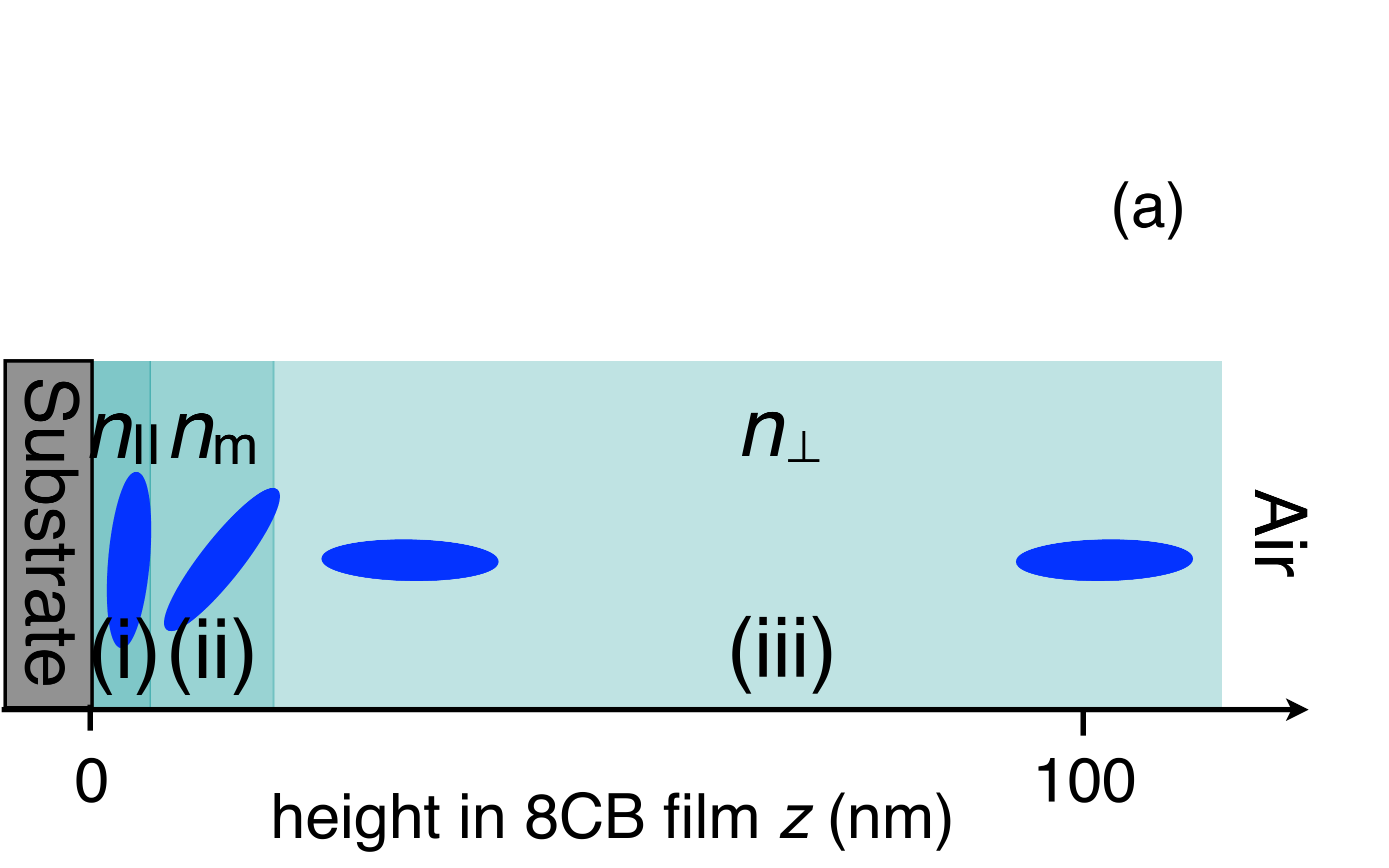}
 \includegraphics[width=2.29in]{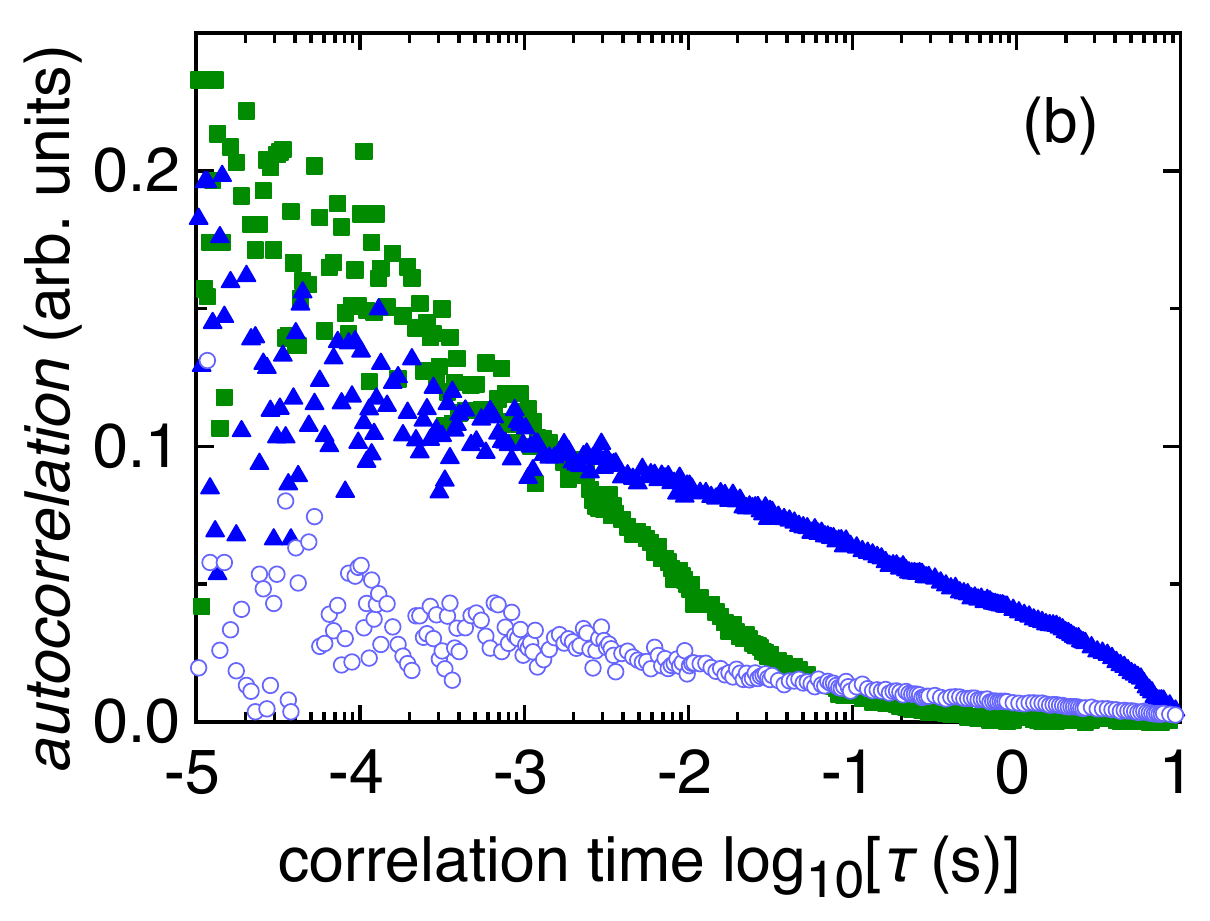}
 \includegraphics[width=2.6in]{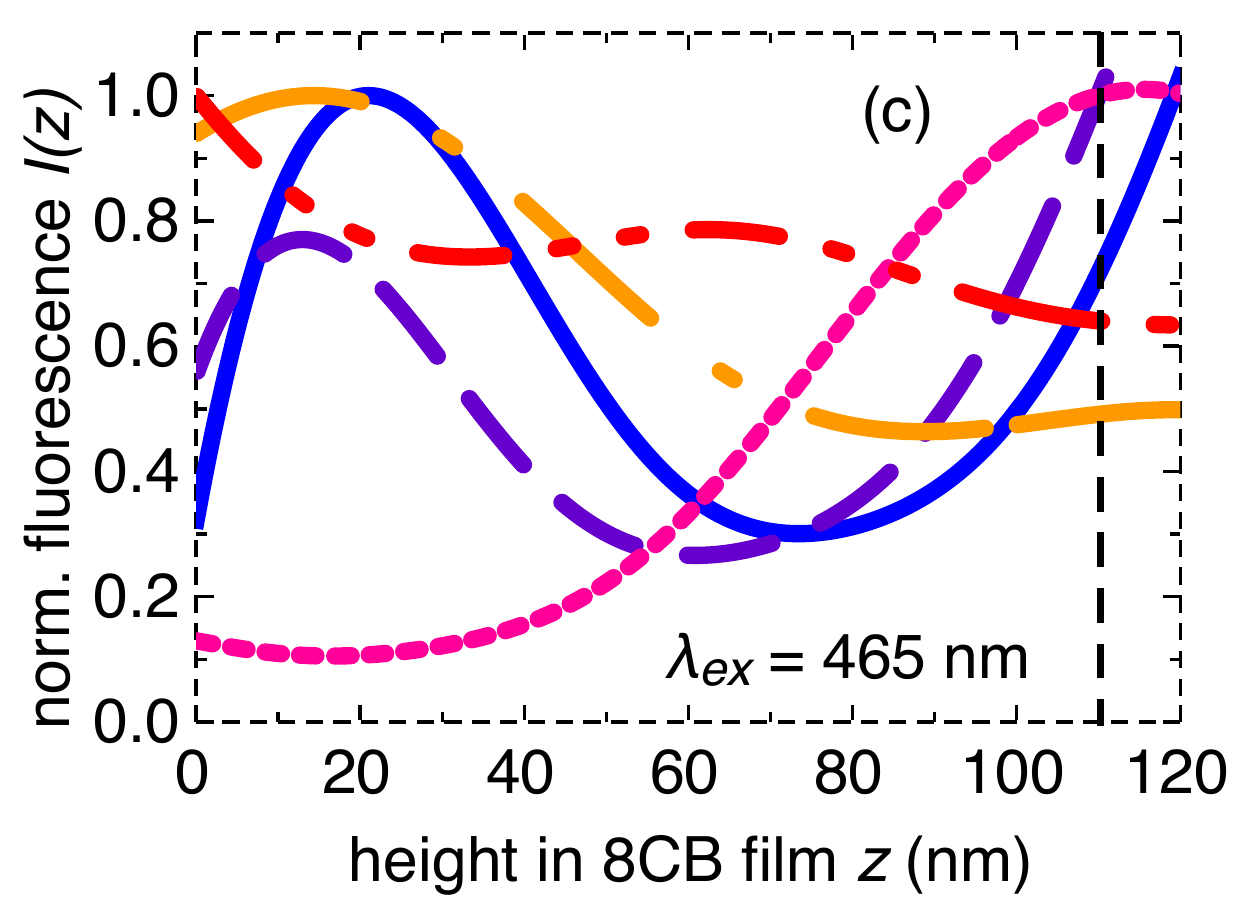}
 \includegraphics[width=2.6in]{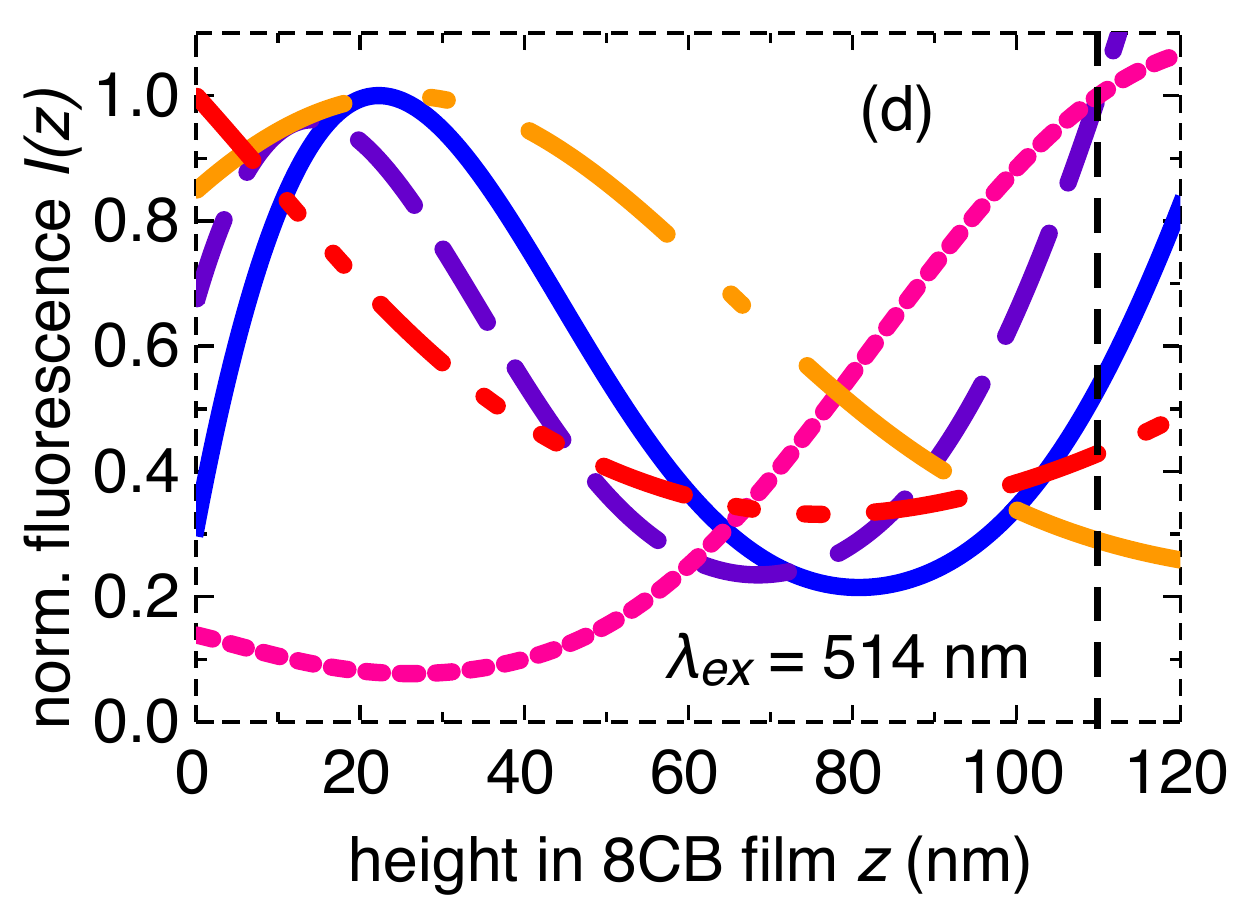}
\caption{\label{interference}a) Scaled model (see FIG.~\ref{structures}) visualizing the vertical LC film structure used in c,d). b) FCS curves obtained with $\lambda_{ex}=465$~nm for PDI in 110~nm thick 8CB films on glass (o) and on silicon wafers with native ($\blacksquare$) and with 100~nm thick thermal oxide ($\blacktriangle$). c,d) Calculated fluorescence intensities $I(z)$ in 8CB films as a function of distance $z$ to the substrate, using Essential Macleod\texttrademark\, for native (blue, solid), 10~nm (purple, dash), 25~nm (pink, dot), 70~nm (yellow, dash-dot), and 100~nm (red, dot-dash-dot) oxide with c) $\lambda_{ex}=465$~nm and d) $\lambda_{ex}=515$~nm.}
\end{figure}
A comparison of FCS curves obtained from 8CB films on silicon wafers to such on glass slides shows a faster component (shorter correlation time) for the FCS curves on the silicon wafers, in particular for the native oxide ($\blacksquare$), see FIG.~\ref{interference} b). Diffusion coefficients of perylenediimides in 8CB have been reported to be in the range of  2 to 5~$\rm\mu m^2/s$ from single molecule tracking experiments and fluorescence recovery experiments at room temperature~\cite{tauber_single_2013}. Neglecting the influence from the interference pattern, a conventional two-component fit of FCS curves obtained on silicon wafers yields a one to two orders of magnitude larger fast component~\cite{tauber_single_2013}. Taking the interference pattern into account, the application of (\ref{zerog},~\ref{firstg}) yields diffusion coefficients similar to those reported from other experiments, as will be detailed below. For FCS curves on glass substrates, the small reflections on the interfaces can be neglected. However, as can be seen in FIG.~\ref{interference} b), the amplitudes of the FCS curves on glass (o) are very low. Conventional FCS is applicable to detect fast diffusion down to the range of 1~$\rm\mu m^2/s$~\cite{woll_polymers_2009, tauber_influence_2013}. Very stable experimental conditions are needed to monitor slowly diffusing molecules passing through the focal volume. At the required long timescales, background fluctuations compete with the signal form the diffusing dye molecule. Consequently, the FCS curve from the 8CB film on glass is on the edge of detectability. 

The situation is different for the films on silicon wafers. There, the vertical fluorescence modulation leads to fluorescence fluctuations on shorter timescales. However, the FCS curves for the films on 100~nm oxide ($\blacktriangle$) and on native oxide ($\blacksquare$) differ significantly. As can be seen from FIG.~\ref{interference} c,d) the interference patterns highlight different regions in the 8CB films on these substrates. For better comparison, in FIG.~\ref{interference} only $I(z)$ calculated for the most reasonable extensions $\zeta$ of the disordered regions are shown, for more details see appendix~\ref{FitsExp}. The films on native (blue, solid line) and on 10~nm thick oxide (purple, dash) were modeled with 3~nm, 12~nm, and $95\pm10$~nm for regions (i), (ii) and (iii), respectively, while the film on 25~nm oxide (pink, dot) contained slightly more disorder having extensions of 3~nm, 22~nm, and $85\pm10$~nm for regions (i), (ii) and (iii), respectively. For the films on 70~nm (yellow, dash-dot) and 100~nm (red, dot-dash-dot) oxide, extensions of 5~nm, 30~nm, and $75\pm10$~nm, respectively, were used. For a detailed discussion of experimental and fit results from the varied combinations of $d$ and $\lambda_{ex}$ we refer to appendix~\ref{FitsExp}. In the following part we now will discuss the implications of varying $d$ on the LC structure at the solid-liquid interface as found from these results.

\subsection{Extension of LC reorientation at the solid-liquid interface\label{LC_disorder}} 
In agreement with previous results~\cite{schulz_influence_2011, tauber_guest_2013} we found a structural difference for the films on thin and on thick oxide. Our results point to a varying thickness $\zeta$ of the interfacial region at the substrate within which the LC molecules reorient from random planar~\cite{mullin_properties_1989} to smectic layering with the LC director normal to the substrate~\cite{designolle_afm_2006}. For thin oxide ($d\leq 10$)~nm, we find $\zeta\approx 15$~nm, while for $d=70$~nm $\zeta\approx 35$~nm. 

The similar surface geometry and roughness for thick and thin oxide~\cite{tauber_influence_2013} induce a similar anchoring geometry of the 8CB at the different substrates. But a reduced mobility at the solid-liquid interface and a higher anchoring strength are expected for small $d$ due to the stronger influence from the underlying silicon in that case. Long range forces inducing fluctuations of the electromagnetic field (as van der Waals forces) are stronger for the more polar silicon substrate than for the less polar silicon oxide~\cite{parsegian_van_2006, israelachvili_intermolecular_2011, loskill_is_2012}. Theoretical considerations~\cite{bocquet_flow_2007} as well as experiments on protein adsorption~\cite{loskill_is_2012} and dynamics in ultrathin liquid films~\cite{tauber_influence_2013} show a decrease in mobility at the solid-liquid interface for an increase of long range van der Waals interactions. In LC materials, due to their optical anisotropy, the LC structure is related to an anisotropy of the van der Waals interactions~\cite{sarlah_van_2001}. Additionally, the so called "Pseudo Casimir" effect, i.~e.~thermal fluctuations of the LC director field, has to be taken into account~\cite{ajdari_pseudo-casimir_1992}. All these interactions will contribute to the anchoring strength of the LC material at the substrate. Because of the LC reorientation in the film induced by the antagonistic boundary conditions, a theoretical calculation of the anchoring strength is not straightforward for the here studied smectic-A 8CB films on silicon substrates. So far, theoretical calculations of the van der Waals interaction and the Pseudo Casimir effect, have only been reported for simple geometries with no reorientation at the walls~\cite{ajdari_pseudo-casimir_1992, sarlah_van_2001}. For nematic LC materials, strong anchoring will stabilize the film, while weak anchoring may lead to the appearance of structural changes and instabilities~\cite{lin_note_2013, effenterre_coupling_2003, guo_influence_2009, de_gennes_physics_2002}. For smectic LC films with antagonistic boundary conditions, melting of the smectic layers close to the substrate into the nematic phase is suggested~\cite{lacaze_bistable_2004, schulz_influence_2011}. Our findings suggest that for thin $d$ such melting exists within a thin region $\zeta\approx 15$~nm, while the remaining part of the film consists of smectic~A layers. 
In contrast, for thick $d$ we found a large extension $\zeta\approx 35$~nm, of the interfacial disorder in the smectic~A 8CB, which matches expectations of an increased instability and structural change due to weaker anchoring found for nematics. For 200 nm thick 8CB films with $d=100$~nm, temperature cycling into the nematic phase was reported to lead to the appearance of structural reorientations, so called focal conic domains (FCD), while for films on native oxide no FCD were found~\cite{schulz_influence_2011}. Although FCD were not found on spin cast films without annealing, their appearance points to a different organization of the 8CB films on thick and thin oxide. Our findings suggest the existence of structural disorder close to the substrate already for as prepared spin cast films on thick oxide.

\subsection{Potentials and diffusion coefficients\label{diff_dyn}} 
\begin{figure}[htb]
 \includegraphics[width=4
in]{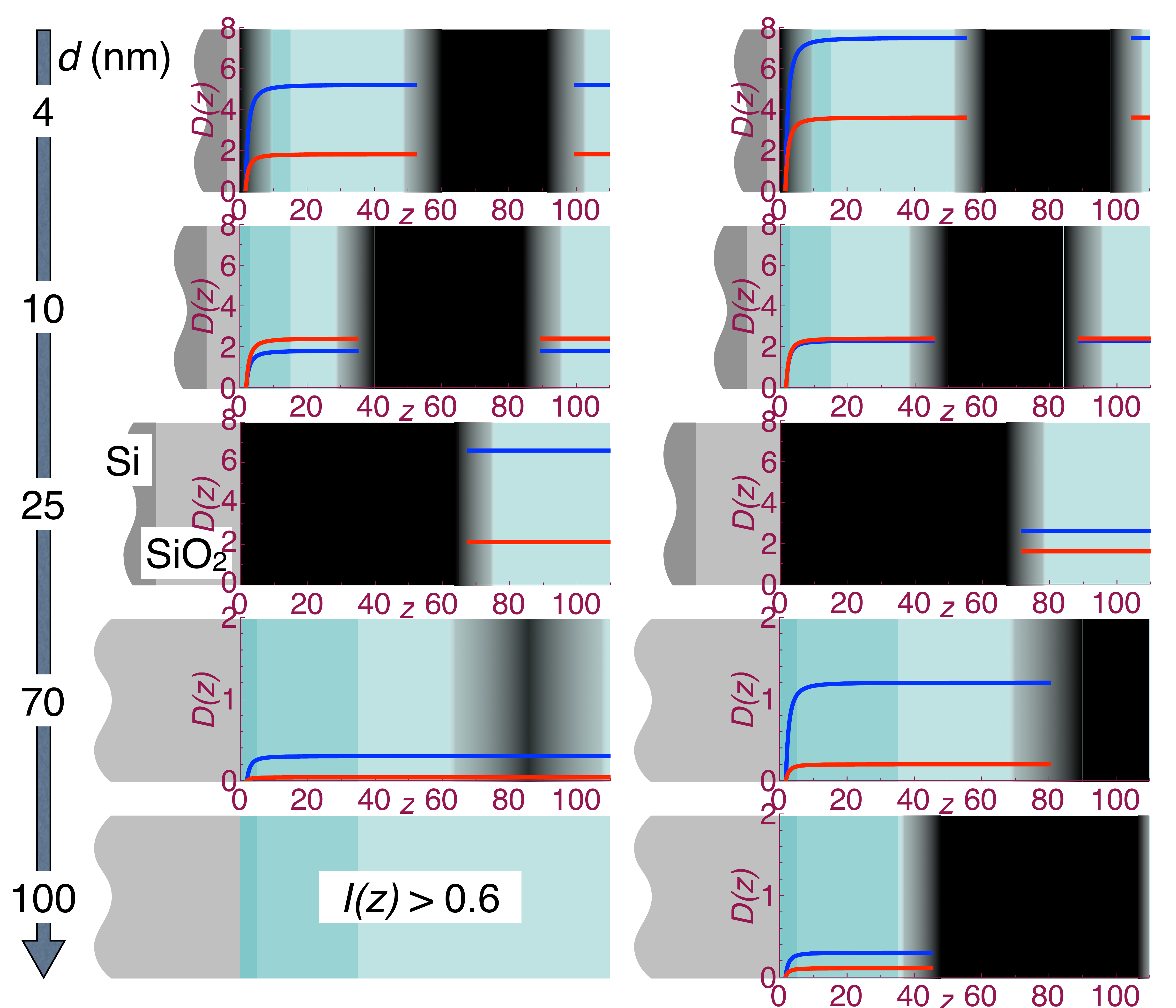}
\caption{\label{DiffRes}Illustration of obtained diffusion coefficients $D_{xy}(z)$ (red) and $D_{z}(z)$ (blue) in $\rm\mu m^2/s$ from each experiment for (left) $\lambda_{ex}=465$~nm and (right) $\lambda_{ex}=515$~nm, and $d$ decreasing from top to bottom. For errors in $D(z)$ see TABLE~\ref{TableDiffusion}. Note the change in scale for thin and thick oxides. Approximations of $I(z)$ are depicted within each 8CB film in greyscale as illustrated in FIG.~\ref{structures}d).}
\end{figure}

Having discussed the LC structure within the 8CB films, we now turn to the diffusion dynamics within these films. An overview of diffusion coefficients $D_{xy}, D_{z}$ and potentials $V_{0}$ derived for the particular $\zeta$, is given in TABLE~\ref{TableDiffusion} together with the calculated vertical area of observation for each experimental configuration. As stated above, the potentials $V(z)$ are modeled to account for vertical modulation of the diffusion coefficients in the area of observation, which not necessarily covers the total film thickness $L$. FIG.~\ref{DiffRes} illustrates $D_{xy}(z)$ (red) and $D_{z}(z)$ (blue) using the results summarized in TABLE~\ref{TableDiffusion}, however, for better visibility we did not plot the determined errors. Along with the vertical dependence of the diffusion coefficients, also the contributing vertical film regions are illustrated by depicting approximations of $I(z)$ as dark and bright regions. We will now summarize and discuss the findings concerning $D_{xy}(z)$ and $D_{z}(z)$ from fitting the experimental data. For a more detailed discussion of the data and fits we refer to appendix~\ref{FitsExp}.

\begin{table}
\caption{\label{TableDiffusion}Diffusion coefficients $D_{xy}, D_{z}$ and potentials $V_{0}$ derived from fits to experimental FCS curves from $110$~nm thick 8CB films on substrates with different oxide thickness $d$, with modeled height $\zeta$ of disordered 8CB at the solid-liquid interface, and indication of vertical region $z$ for which $I(z)\gtrsim0.5$ and which thus contributes to the derived diffusion coefficients from fitting. Errors denote standard deviations from multiple experimental data.}
\begin{ruledtabular}
\begin{tabular}{lcccccc}
$d$ $[\rm nm]$ & $\lambda_{ex}$ [nm] & $D_{xy}$ $[\rm\mu m^2/s]$ & $D_{z}$ $[\rm\mu m^2/s]$ & $V_{0}$ $[\rm meV]$ & $\zeta$ $[\rm nm]$&region $z$ [nm]\\
\hline
4 & 465 & $1.8\pm0.3$ & $5.2\pm2.6$ & $-0.006\pm0.0004$ & 15&$5<z<50$, $z>100$\\
4 & 515 & $3.6\pm0.9$ & $7.5\pm2.9$ & $-0.006\pm0.001$ & 15&$5<z<50$, $z>105$\\
10 & $465^*$ & $2.4\pm0.6$ & $1.8\pm0.9$ & $-0.007\pm0.003$& 15& $z<35$, $z>90$\\
10 & 515 & $2.4\pm0.3$ & $2.3\pm0.5$ & $-0.006\pm0.001$ & 15& $z<45$, $z>90$\\
25 & 465 & $2.1\pm0.5$ & $6.6\pm3.2$ & 0 & 25&$z>70$\\
25 & 515 & $1.6\pm0.3$ & $2.6\pm0.9$ & 0 & 25&$z>70$\\
70 & 465 & $0.04\pm0.02$ & $0.3\pm0.1$ & $-0.007\pm0.0005$ & 35&$z<70$\\
70 & 515 & $0.2\pm0.1$ & $1.2\pm0.7$ & $-0.004\pm0.001$ & 35&$z<80$\\
100 & 465 & $0.03\pm0.03$ &  $0.2\pm0.2$ & $-0.01\pm0.001$ & 35&$0\leq z\leq L$\\
100 & 514 & $0.11\pm0.06$ & $0.3\pm0.2$ & $-0.006\pm0.001$ & 35&$z<40$\\
 \end{tabular}
\end{ruledtabular}
{\footnotesize{* values obtained for film thickness $L=120$~nm}}
\end{table}

For $d=25$~nm, the fluorescence signal is detected from $z>70$~nm, see the depiction in FIG.~\ref{DiffRes} middle row. Within this region, the 8CB is organized in smectic-A layers and a vertical modulation of diffusion coefficients should be negligible, in agreement with $V_0=0$ obtained from fitting. The derived diffusion coefficients (see TABLE~\ref{TableDiffusion}) are within errors similar to the anisotropic diffusion coefficients $D_{\perp}=2.4\pm0.1$~$\rm\mu m^2/s$ and $D_{\parallel}=3.8\pm0.6$~$\rm\mu m^2/s$ reported for PDI in 200~nm thick 8CB films from single molecule tracking (SMT) experiments at room temperature~\cite{schulz_influence_2011, tauber_single_2013}. From SMT of a perylenediimide in a 0.5~$\rm\mu m$ thick 8CB cell, anisotropic diffusion coefficients were reported as $D_\perp=2.8\pm0.5$~$\rm\mu m^2/s$ and $D_\parallel=4.4\pm0.5$~$\rm\mu m^2/s$~\cite{pumpa_slow_2012}. In the thicker cell with parallel alignment at the walls, interfacial effects contribute less to the diffusion coefficients, explaining slightly larger values than those obtained from our FCS experiment. Both SMT experiments were conducted without temperature control, and a slightly larger $D$ may also be explained by the high temperature sensitivity of 8CB .

Diffusion coefficients obtained from the films on native oxide are of similar range as those for 25~nm oxide, see TABLE~\ref{TableDiffusion}. On native oxide $I(z)$ contains a high amplitude in a narrow upper part of the film (for $z\gtrsim100$~nm) and in the lower part $5<z<50$~nm (depiction in FIG.~\ref{DiffRes} top row), whereby the fluorescence signal from the close proximity to the interface is weak, see FIG.~\ref{4nm}~a) in the appendix~\ref{FitsExp}. As discussed above, $\zeta\approx15$~nm for films on native oxide. Thus, contributions from reorientation are small and most of the signal is recorded from smectic-A layered regions. Therefore, we again can assign the obtained diffusion coefficients to the anisotropic diffusion in 8CB as $D_{xy}=D_{\perp}$ and $D_{z}=D_{\parallel}$. A small attractive potential $V_0=-0.006$~meV is obtained from fitting, taking into account the decrease of the mobility induced by the solid-liquid interface. 

For the remaining three substrates $I(0)\geq0,5$, see FIG.~\ref{interference}, and there is a considerable contribution from the area of reorienting 8CB to the detected fluorescence. As a consequence, $D_{xy}$ and $D_{z}$ can no longer be assigned to diffusion perpendicular and parallel to the LC director. As explained above, within these three substrates we found the extension $\zeta$ of the reorientation region to be smallest for $d=10$~nm. For films on this kind of substrate, the fluorescence signal contains contributions from the lower film region ($z\lesssim40$~nm) and a small upper film region ($z>90$~nm), see FIG.~\ref{DiffRes} second row. The obtained small potential is similar to that for films on native oxide, and the obtained $D_{xy}\approx2.4$~$\rm\mu m^2/s$ is similar to that for $d=25$~nm, while $D_{z}$ is of similar range and thus smaller than for $d=25$~nm. This may be explained by different effects of the interfacial LC reorientation on diffusion parallel and perpendicular to the substrate. Any deviation from the smectic~A layering with its LC director in $z$-direction will lead to a decrease of $D_{z}$ additional to the hydrodynamic effect represented by the interface potential $V_0$.  On the other hand, if there are extended areas of similar orientation with LC director deviating considerably from the substrate normal, $D_{xy}$ may even be enhanced. Since $D_{xy}$ is similar to that for $d=25$~nm, we expect such areas to be small, in agreement with our modeling.

The situation is different for $d\geq70$~nm, see FIG.~\ref{DiffRes} two bottom rows. As can be seen in TABLE~\ref{TableDiffusion}, films on those substrates yielded considerably smaller diffusion coefficients. Thereby, the largest values $D_{xy}\approx0.2$~$\rm\mu m^2/s $ and $D_{z}\approx1.2$~$\rm\mu m^2/s$ were obtained for $d=70$~nm and $\lambda_{ex}=514$~nm (FIG.~\ref{DiffRes} second row from bottom, right). In that case, $I(0)\approx0.85$ is only slightly larger than $I(0)\approx0.7$ for $d=10$~nm, compare FIG.~\ref{10nm} a) and~\ref{70nm} a) in appendix~\ref{FitsExp}. In both cases, $I(z)$ shows large amplitude in the lower film region, containing the reorientation of 8CB. For $d=70$~nm the fluorescence is also detected from the middle of the film, decreasing towards upper film regions, while for $d=10$~nm, the amplitude of $I(z)$ is high again in the upper film region but low in the middle of the film. The attractive potential $V_0=-0.004\pm0.001$~meV is only two thirds of the potential obtained for $d=10$~nm. This agrees with the smaller area of observation, which covers two thirds of the film thickness only. As can be seen in FIG.~\ref{structures} e,f), hydrodynamic considerations according to (\ref{Dpll_noslip},~\ref{Dperp_noslip}) still yield a decrease of diffusion coefficients at $z=80$~nm, which is most pronounced for $D_{\parallel}$ in $z$-direction. This effect can be taken into account for the smaller $D_{z}=1.2\pm0.7$~$\rm\mu m^2/s$ obtained for $d=70$~nm and $\lambda_{ex}=514$~nm compared to $D_{z}=2.3\pm0.5$~$\rm\mu m^2/s$ for $d=10$~nm (see TABLE~\ref{TableDiffusion}). Additionally, the suggested larger extension $\zeta\approx35$~nm for $d=70$~nm compared to $\zeta\approx15$~nm for $d=10$~nm will contribute to the smaller $D_{z}$, since $D_{z}$ is highest within smectic~A layered regions with LC director perpendicular to the substrate. 

In contrast to $D_{z}$ being still of similar order of magnitude, $D_{xy}=0.2\pm0.1$~$\rm\mu m^2/s$ is one order of magnitude smaller than $D_{xy}=2.4\pm0.3$~$\rm\mu m^2/s$ found for $d=10$~nm. As discussed above, LC reorientation may enhance $D_{xy}$ if it leads to extended areas of similar orientation with LC director deviating considerably from the substrate normal. For $d=10$~nm we had found $D_{xy}$ to be similar to bulk $D_{\perp}$, whereby we concluded that such areas had to be small in that case. The decrease of $D_{xy}$ found for $d=70$~nm suggests the absence of areas with similar oriented LC director on the experimental length scale in the range of 100~nm, in agreement with the proposed increased interfacial disorder in the above section. The PDI tracer molecules are known to cause a considerable distortion of the local 8CB structure~\cite{tauber_single_2013}, which in SMT experiments on 200~nm thick 8CB films led to smaller diffusion coefficients than expected from the comparison with self-diffusion coefficients obtained by NMR~\cite{schulz_optical_2010, tauber_single_2013}. In the disordered region close to the substrate, the rearrangement of the surrounding 8CB may be even more pronounced accounting for the observed strong decrease of the lateral tracer mobility.

An even stronger decrease of diffusion coefficients is found for $d=70$~nm with $\lambda_{ex}=465$~nm (FIG.~\ref{DiffRes} second row from bottom, left) and $d=100$~nm with $\lambda_{ex}=514$~nm (FIG.~\ref{DiffRes} bottom row, right), yielding one order of magnitude smaller $D_{z}$, while the $D_{xy}$ are decreased by one to two orders of magnitude, see TABLE~\ref{TableDiffusion}. In that case, $I(0)\approx1$, decreasing to 0.5 within 70 and 40~nm from the substrate, respectively. According to the hydrodynamic model (\ref{Dpll_noslip},~\ref{Dperp_noslip}), at  70 and 40~nm distance to the substrate there is only a small decrease of $D_i$. Hydrodynamic modeling according to (\ref{Dpll_noslip},~\ref{Dperp_noslip}), thus, cannot account for the observed strong mobility decrease. However, the signal is recorded from the disordered region with little contribution from upper film regions with smectic~A layering. As explained above, the extended rearrangement of 8CB around the tracer molecules may cause a strong decrease in mobility. It is now seen to slow both lateral and vertical diffusion, with still a stronger effect on lateral diffusion. For both films, potentials similar to those from films on thin oxide were obtained, see TABLE~\ref{TableDiffusion}, although here the fluorescence is recorded from the lower film region mainly, pointing to a stronger vertical modulation of the diffusion compared to that for $d\leq25$~nm. 

As mentioned above, for $d=100$~nm with $\lambda_{ex}=465$~nm, $I(z)\geq0.6$ in the entire film, causing only weak intensities of fluorescence fluctuations from vertical diffusion, which in turn leads to a small amplitude of the FCS curve (see FIG.~\ref{interference} b) rendering further discussion of the obtained diffusion coefficients not feasible.

In general, stronger long-range interactions with the substrate are expected to cause a stronger decrease of the mobility of the liquid at the interface~\cite{tauber_influence_2013, loskill_is_2012, bocquet_flow_2007}. For the here studied smectic~A films we find a stronger decrease of the tracer mobility for the films which are subject to weaker long-range interactions with the substrate. However, the observations made for simple liquids may not hold in the same way for complex liquids as the here studied smectic~A 8CB. For the latter, a slowdown of tracer diffusion is reported in respect to hydrodynamic considerations based on diffusion coefficients for self-diffusion~\cite{tauber_single_2013}, which is explained by a necessary rearrangement of the LC around the diffusing tracers. As mentioned in the previous section~\ref{LC_disorder}, in 8CB films on 100~nm thick oxide, FCDs representing structural rearrangement were found after annealing, while they were not observed on native oxide~\cite{schulz_influence_2011}. Our findings suggest that already in the as prepared films on thick oxide irregularly sized LC domains exist. LC materials are very sensitive to surface structure~\cite{roscioni_predicting_2013, coursault_self-organized_2015}. It is therefore reasonable to assume that the grain boundaries are fixed by some surface structure, which could be given by irregularly distributed surface silanols known to be present on silicon oxide grown on silicon wafers~\cite{albers_driven_2015}. On the other hand, transition of grain boundaries by the tracer molecules implies their rearrangement, which would cause a slowdown or even repulsion of the tracers. 

The absence of FCDs observed for 8CB on native oxide~\cite{schulz_influence_2011}, points to a different internal structure of these films on thin oxides, in agreement with our observations. For thin oxides, the underlying silicon enhances the long-range interactions~\cite{tauber_influence_2013, loskill_is_2012, bocquet_flow_2007}, thus leading to even stronger anchoring of 8CB at the substrate. FCDs as well as dislocations reported for 8CB on rubbed PVA contain regions with unfavorable perpendicular orientation of 8CB at the substrate~\cite{designolle_afm_2006, coursault_self-organized_2015}, which seem to be absent in thin 8CB films on thin oxide. Instead, for native oxide, nematic melting of 8CB at the substrate is suggested~\cite{lacaze_bistable_2004, schulz_influence_2011}. In the nematic phase, the self-diffusion is enhanced in respect to the smectic phase~\cite{dvinskikh_anisotropic_2002}, which can counterbalance a slightly reduced mobility due to stronger long-range interactions with the substrate. Moreover, the absence of grain boundaries for 8CB on thin oxides facilitates the tracer diffusion in respect to 8CB films on thick oxides and thus explains the agreement of the obtained diffusion coefficients with the hydrodynamic model.
\vspace{2cm}

\section{Conclusion}
The here developed correlation function (\ref{zerog},~\ref{firstg}) could be successfully applied to determine the tracer diffusion behavior of a perylenediimide in thin 8CB films on silicon substrates with varied oxide thickness. In particular, the fit results allowed us to estimate the extension $\zeta$ of the interfacial reorientation of the 8CB close to the substrate from three different modeled extensions. We found $\zeta$ increasing with oxide thickness, which we assign to a decreasing stabilization of the liquid crystal structure by long-range interactions with the underlying silicon in accordance with literature~\cite{tauber_influence_2013, loskill_is_2012, bocquet_flow_2007, ajdari_pseudo-casimir_1992, sarlah_van_2001, effenterre_coupling_2003, lin_note_2013}. For thin oxides ($d\leq10$~nm), $\zeta\approx15$~nm was obtained, while for thick oxides ($d\geq70$~nm), a larger disordered region with $\zeta\approx35$~nm was found. For 200~nm thick 8CB films different LC structures after annealing into the nematic phase are reported in literature for native and for 100~nm thick oxides~\cite{schulz_influence_2011}. Our results suggest a different structure for the spin cast films even before annealing.

Our findings not only show a variation in the structural arrangement of the 8CB at the substrate. The larger extend of re-orientation for the thick oxides also leads to a stronger decrease of tracer diffusion close to the substrate. Thereby, $D_z$ is reduced about one order of magnitude (within $\approx 50$~nm from the substrate), while $D_{xy}$ is reduced even more, about two orders of magnitude. This strong decrease of tracer diffusion can be explained by dislocations and grain boundaries~\cite{coursault_self-organized_2015} in the disordered region, which slow or even repel diffusing tracers by hampering the local rearrangement caused by these tracers~\cite{tauber_single_2013}. In the upper film regions on the thin oxide, anisotropic diffusion coefficients  of similar order than reported from SMT experiments~\cite{schulz_optical_2010, pumpa_slow_2012, tauber_single_2013} were found, in particular $D_{\perp}=2.1\pm0.5$~$\rm\mu m^2/s$ and $D_{\parallel}=6.6\pm3.2$~$\rm\mu m^2/s$ from investigation of films on 25~nm thick oxide with excitation at 465~nm, where $\approx40$~nm of the upper film region contributed to the signal. The very narrow height ($\leq30$~nm) of this region in case of excitation at 514~nm led to considerable errors from fitting going along with diffusion coefficients of half of the expected size, which we assign to an influence from higher orders of the correlation function. Thus, for very thin film regions contributing to~(\ref{zerog},~\ref{firstg}), higher orders have to be taken into account.

The fluorescence modulation at reflecting substrates may further be applied to investigate thin polymer films, in particular to yield additional information about the widely discussed impact from confinement effects to the glass transition temperature~\cite{baumchen_reduced_2012, krutyeva_effect_2013}. Investigations of spectral diffusion~\cite{krause_freezing_2011, araoz_cage_2012} or rotational diffusion~\cite{adhikari_temperature_2011, araoz_influence_2014} could be studied for different vertical regions within such films without the need to change the film thickness.

\appendix
\section{\label{Experiment}Experiment}
We performed FCS on liquid crystal (LC) films on silicon wafers with native ($\approx 4$~nm~\cite{loskill_is_2012}), and thermally grown $10$~nm, $25$~nm, $70$~nm (all: Center for Microtechnologies, Chemnitz), and $100$~nm oxide (CrysTec, Berlin), employing two different excitation wavelengths ($\lambda_{ex}=465$~nm and $\lambda_{ex}=515$~nm). The silicon wafers were slightly doped (resistivity $5-20$~$\rm\Omega cm$). Glass substrates (Carl Roth, Karlsruhe) were used for further comparison. Liquid crystal 4-n-octyl-4'-cyanobiphenyl (8CB, SYNTHON Chemicals, Wolfen) is in its smectic-A phase at room temperature and was used without further purification. The 8CB films were prepared as previously described~\cite{schulz_optical_2010} by spin coating solutions of $50$~mg/ml 8CB in Toulene, resulting in film thicknesses $L=110\pm10$~nm. The solutions had been doped with N,N'-di-propyl-1,6,7,12-tetra-(4-heptyl-phenoxy)-perylene-3,4,9,10-tetra-carboxdiimide (PDI), a perylenediimide dye synthesized by Melanie Bibrach, TU Chemnitz, yielding nanomolar concentration of PDI in the 8CB films. Perylenediimides are very photostable dyes suited for single molecule investigation. The here chosen PDI does not align with the LC structure, but will rotate freely within the LC film, yielding an isotropic fluorescence signal on our experimental time scale with correlations times $\tau\geq10\,\rm\mu$s~\cite{schulz_optical_2010}. Chemical structures of 8CB and PDI are given in FIG.~\ref{structures} a) and b). 

To aid focussing, highly diluted Nile Red doped Latex Beads of $0.2$~$\rm\mu m$ diameter (Molecular Probes, Leiden, Netherlands) were spin coated on the clean substrates from solution in water before the 8CB film preparation. 
Defocussing of the confocal excitation on the sample in the range of $\rm\Delta z>0.1$~$\rm\mu m$ had lead to considerable drops in the amplitude of the recorded fluorescence in combination with apparent slower diffusion, as reported by Petrov and Schwille in~\cite{resch-genger_state_2008}. By aid of the Latex beads, $\rm\Delta z$ cold be kept in a range below $0.1$~$\rm\mu m$. 

FCS data were obtained using a previously described homebuilt confocal microscope~\cite{schulz_optical_2010, schmidt_change_2012} with an avalanche photodiode (SPCM-AQRH-14, Excelitas), a TCSPC-board (Becker \& Hickl SPC-630) and a microscope objective with high numerical aperture $N_{\rm A}=0.9$. Since the self-diffusion coefficient of 8CB is very temperature-sensitive~\cite{dvinskikh_anisotropic_2002}, a water cooling device together with a pt100 sensor was implemented into the sample stage. This allowed for stabilization at $T=22\pm0.3^\circ$C with a calibration accuracy of $\pm1^\circ$C. The setup was eventually extended by implementing a second excitation beam at $515$~nm employing a 25 mW cw laser (Cobolt Fandango). Circular excitation at the sample stage was obtained using sets of $\lambda/2$ and $\lambda/4$ plates in each excitation beam~\cite{tauber_characterization_2011}. The  excitation light was expanded by a linear beam expander (Thorlabs) to fill the back aperture of the objective.
To avoid photobleaching, the excitation power was dimmed to 1 $\rm\mu W$ (cw equivalent) for both excitation wavelengths. Before recording FCS data, a region of $100$~$\rm\mu m$~x~$100$~$\rm\mu m$ on the sample was scanned in reflected excitation light (by removing the long pass filter in the detection beam). Regions showing structural artifacts from preparation could thus be excluded from recording. For FCS recordings also a distance of at least $2$~$\rm\mu m$ was kept to Latex beads to avoid the influence from structural rearrangements of the liquid crystal close to the beads~\cite{koenig_characterization_2009, tauber_single_2013}. For each of the ten combinations of excitation wavelength and substrate a fresh sample was prepared and 10 to 15 FCS measurements were taken at different positions on the sample, with an acquisition time of 
15~minutes for each one. All experimental FCS data sets were fitted separately. For $\lambda_{ex}=465$~nm on the 70~nm oxide, several data sets contained a rather low amplitude of the autocorrelation curve, pointing to some higher experimental instability during measurements. These data sets were excluded from further evaluation, leaving only seven valid data sets. In general, the FCS curves were less noisy for excitation at 514~nm, which is due to the stronger absorption of PDI at that wavelength. 

The focal width $\omega_{xy}$ of the exciting laser beam was determined by an approximation for diffraction limited focussing~\cite{kawai_anisotropic_2004} 
\begin{equation}
w_{xy} = \frac{2\lambda_{ex}}{\pi n {A_N}}\, ,
\label{FCSwidth}
\end{equation} 
where $n$ is the mean refractive index of 8CB at $22^\circ$C and $A_N$ is the numerical aperture of the used objective. We obtained $w_{xy}=0.21$~$\rm\mu m$ and $w_{xy}=0.23$~$\rm\mu m$ for $\lambda_{ex}=465$~nm and $\lambda_{ex}=514$~nm, respectively.

\section{\label{DerivCorr}Derivation of density correlation function}
For calculating the density-density correlator entering eq.~(\ref{start}) we used a mathematical model of diffusion in a film of certain thickness with inhomogeneous properties of the media along the transverse $z$ and homogeneous ones along the lateral direction. An analytical solution of the diffusion equation is not feasible. As explained in section~\ref{FCSfunction}, the $z$-dependence of the diffusion coefficients is induced by the interface with the substrate and decays rapidly with distance to the substrate. Thus, for a sufficiently thick film, the $z$-dependence of the diffusion coefficients is only weak.  We use this thick-film assumption for deriving approximated relations for the ACF stated in eq.~(\ref{start}). In more details the model is formulated as follows:
\begin{itemize}

\item[(i)] The volume $G$ in eq.~(\ref{start}) is the volume of the film of thickness $L$, $0\le z\le L$ infinitely extended in $x$ and $y$ directions.

\item[(ii)] The properties of the film are isotropic and uniform in the $x,y$-plane. Therefore, the dependence of the diffusion coefficients  can be reduced to dependence on the $z$-coordinate only. Thus the diffusion tensor has only two components, $D_{xy}(z)$ and $D_z(z)$.
 
\item[(iii)] Since the tracers are highly diluted in the film, the density-density correlator $\langle\rho(\mathbf{r},t)\rho(\mathbf{r}',t')\rangle$ of the active particles can be replaced by their joint probability density, which, in turn, can be represented as a product of the conditional probability, $P(\mathbf{r},t |\mathbf{r}^{\prime},t^{\prime})$, to find the particle at time $t$ in the small volume positioned at $\mathbf{r}$ if it was in $\mathbf{r}^{\prime }$ at time $t^{\prime}$ and of some initial distribution of particles. Ergodicity and stationarity of the diffusion process allows to assume that the correlators in~(\ref{start}) depend only on the time interval $\tau=t-t^{\prime}$ and can be described by the Fokker-Planck equation of the form
\begin{equation}
\frac{\partial P}{\partial \tau }=\frac{1}{2}D_{xy}(z)\left(\frac{\partial ^{2}}{\partial x^{2}}+\frac{\partial ^{2}}{\partial y^{2}}\right)P+\frac{1}{2}\frac{\partial ^{2}}{\partial z^{2}}D_{z}(z)P-\frac{\partial }{\partial z}a(z)P,
\label{FPG1}
\label{Fokker2}
\end{equation}
supplemented by the initial condition $P(\mathbf{r},0 |\mathbf{r}^{\prime},0)=\delta(\mathbf{r}-\mathbf{r}^{\prime})$.

\item[(iv)] There is no flux on the interface of the film. Thus the boundary conditions have the form
$$
\left[a(z)P-\frac{1}{2}\frac{\partial}{\partial z}D_{z}(z)P\right]_{z=0,L}=0.
$$

\item[(v)] The stationary probability distribution, i.e. the conditional probability $P(\mathbf{r},\tau |\mathbf{r}^{\prime},0)$ in the limit $\tau\to\infty$, depends only on $z$ and is assumed to satisfy the Gibbs-distribution in the potential $V(z)$ (see discussion in section III),
$$
P(\mathbf{r},\tau\to\infty |\mathbf{r}^{\prime},0)=N_0^{-1}\exp\left[-\frac{V(z)}{k_B T}\right],\qquad N_0\equiv\int_0^Ldz \exp\left[-\frac{V(z)}{k_B T}\right].
$$
The latter being substituted into~(\ref{Fokker2}) gives rise to an expression for the drift term
$$
a(z)=\frac{1}{2}D_z(z)\frac{\partial}{\partial z}\left[\ln D_z(z)-\frac{V(z)}{k_B T}\right].
$$

\item[(vi)] Since the potential $V(z)$ is generated by the film interfaces, we can assume that in case of a thick film it is weak. The same assumption one can make for the diffusion coefficients as well and represent them as  $D_z(z)=D^{(0)}_z+\delta D_z(z)$ and $D_{xy}(z)=D^{(0)}_{xy}+\delta D_{xy}(z)$.   
\end{itemize}

The last assumption allows to solve equation~(\ref{Fokker2}) perturbatively. By the method of separation of variables we derive the following expression for the leading order of $g(\tau)$:
\begin{eqnarray}\label{zerog}
g^{(0)}(\tau ) &=& \int\int_Gd^3rd^3r'\; \frac{ I_0^2I(z)I(z') }{2\pi L^2 D^{(0)}_{xy}\tau} \exp \left\{ -\frac{x^{2}+y^{2}}{2w_{xy}^{2}} -\frac{x^{\prime 2}+y^{\prime 2}}{2w_{xy}^{2}}\right\}  \nonumber\\
&\times&\exp \left\{ -\frac{1}{2D_{xy}^{(0)}\tau }\left[ \left( x-x^{\prime
}\right) ^{2}+(y-y^{\prime })^{2}\right] \right\} 
\sum \limits_{n=0}^{\infty} {\rm e}^{
- \frac{n^2 \pi^2 D_{z}^{(0)}\tau}{2L^2}
} \cos \frac{n\pi z^{\prime}}{L}
\cos \frac{n\pi z}{L}\nonumber\\
&=&\frac{I_0^2}{1+D_{z }^{(0)}\tau/(4w_{xy}^2)}
\sum \limits_{n=0}^{\infty}  
{\rm e}^{- \frac{n^2 \pi^2 D_{xy}^{(0)}\tau}{2L^2}} a_n^2
\end{eqnarray}
where
\begin{equation}
a_n = \frac{1}{L}\int\limits_{0}^{L} d\xi \cos \frac{n\pi \xi}{L} I(\xi)\, .
\end{equation}

Employing a Fourier-transformation in $x$ and $y$ coordinates and a Laplace transformation in $\tau$, the first order correction can be obtained as well: 
\begin{eqnarray}\label{g1}
g^{(1)}(\tau) &=&
\frac{I_0^2}{1+D_{z }^{(0)}\tau/(4w_{xy}^2)}
\left [
\frac{2}{\pi}
\sum \limits_{n,m=0}^\infty n 
V_{nm} \phi_{nm} \left( 
\frac{D_{xy}^{(0)}\tau}{2L^2}
 \right) a_n a_m \right .
\nonumber \\
&+&
\left . \frac{1}{2}\sum \limits_{n=0}^{\infty} 
{\rm e}^{- \frac{n^2 \pi^2 D_{xy}^{(0)}\tau}{2L^2}}
a_n \left [ a_n v_0
-\sum \limits_{k=1}^\infty v_k
\left( a_{n+k}+a_{n-k} \right) \right ] \right ] \, , 
\label{firstg}
\end{eqnarray}
where
\begin{equation}
\phi_{nm} (\zeta) = \frac{
{\rm e}^{-m^2\pi^2 \zeta}-
{\rm e}^{-n^2\pi^2 \zeta}}{n^2-m^2}.
\end{equation}
The Fourier transform $V_{nm}(a,b)$ of the potential $V(z)$ is given by
\begin{equation}\label{VV}
V_{nm}=\int\limits_{0}^{L} \frac{d\xi}{ k_{B}T}V^{\prime }(\xi )
\sin \frac{n\pi \xi}{L} \cos \frac{m\pi \xi}{L}
\, ,\quad\mbox{and}\quad
v_k=\frac{1}{Lk_BT} \int \limits_0^L d\xi V(\xi) \cos \frac{\pi k \xi}{L}\, .
\end{equation}  

The zero order term~(\ref{zerog}) and first order term~(\ref{firstg}) now can be used together with numerical data for $I(z)$ to derive lateral and vertical diffusion coefficients $D_{xy}$ and $D_z$ as well as the strength $V_0$ of the decaying potential for the tracers in the films of a given thickness $L$.

\section{\label{hydro}Height dependent diffusion coefficients according to hydrodynamic model}
As stated above, the derived correlation function will only take vertical dynamics into account, which occur within the area of observation determined by the vertical fluorescence modulation. We therefore compared the results with expectations using continuum hydrodynamics. Modeling confined diffusion is not straightforward, and analytical solutions for the diffusion coefficients derived from hydrodynamic considerations exist only for few cases~\cite{bickel_note_2007}. In general, the proximity of a solid-liquid interface will affect the self-diffusion behavior of a liquid, and thus the diffusion of incorporated tracer molecules. Previous experimental reports showed a slowdown of the self-diffusion of 8CB at the walls of silica nanopores~\cite{lefort_relation_2008} as well as of the tracer diffusion of perylenediimide in thin 8CB films on silica~\cite{schulz_optical_2010}. For the room temperature nematic homolog 5CB, conservation of the bulk diffusion anisotropy at the solid-nematic interface was reported~\cite{chakraborty_anisotropic_2015}. Joly et al.~discussed the application of models from continuum hydrodynamics to nanohydrodynamics. According to their results, for nanohydrodynamics on wetting surfaces, a hydrodynamic model with no slip boundary condition may be applied~\cite{joly_probing_2006}. Lin et al.~give approximations for the diffusion parallel to the substrate $D_{xy}(z)$ and perpendicular $D_z(z)$ to it, derived from a simple hydrodynamic model employing hard spheres of diameter $a$ and bulk diffusion coefficient $D$ at a solid wall with no slip boundary condition~\cite{lin_direct_2000}

\begin{equation}
\frac{D_{xy}(z)}{D}\approx 1-\frac{9}{16}\left(\frac{a}{z}\right)+\mathcal{O}\left(\frac{a}{z}\right)^3\!,
\label{Dpll_noslip}
\end{equation}
\begin{equation}
\frac{D_z(z)}{D}\approx 1-\frac{9}{8}\left(\frac{a}{z}\right)+\mathcal{O}\left(\frac{a}{z}\right)^3\!.
\label{Dperp_noslip}
\end{equation}

Previous experiments have shown that in the here used liquid crystal 8CB the dye molecules lead to a large re-orientation of the surrounding LC material~\cite{tauber_single_2013}, which can be modeled by an increased hydrodynamic radius~$a$~\cite{tauber_single_2013, schulz_optical_2010}. In case of PDI diffusion in 8CB, $a$ was found to be approximately three times the molecular radius~\cite{schulz_optical_2010}, yielding $a=3.5$~nm.  Due to the diffusion anisotropy parallel and perpendicular to the LC director in 8CB, the local diffusion coefficient depends on the particular orientation of the 8CB in the different regions of the film, as was found for diffusion of CdSe quantum dots in a nematic LC cell~\cite{lee_position-dependent_2015}. FIG.~\ref{structures} e,f) shows $D_{xy}(z)$ calculated from~(\ref{Dpll_noslip}) and $D_{z}(z)$ calculated from~(\ref{Dperp_noslip}) for three different bulk diffusion coefficients obtained for perylenediimide in 8CB, (i) diffusion parallel to the liquid crystal director with bulk $D_{\parallel}=4.4$~$\rm\mu m^2/s$~\cite{pumpa_slow_2012, tauber_single_2013} (dash), (ii) diffusion perpendicular to the liquid crystal director with bulk $D_{\perp}=2.9$~$\rm\mu m^2/s$~\cite{pumpa_slow_2012, tauber_single_2013} (dot), and (iii) a mean diffusion coefficient $\bar{D}=3.2$~$\rm\mu m^2/s$~\cite{pumpa_slow_2012, tauber_single_2013} (solid). The here studied thin 8CB films contain structural reorientations close to the substrate. In that area we expect to find deviations from the model.

\section{\label{FitsExp}Fits to experimental FCS curves}
Here we will start the discussion with 8CB films on 10 and on 70~nm oxide, where $I(z)$ highlights the disordered regions. Then we will turn to 25~nm oxide, where the fluorescence is detected from the upper film region only. On native oxide, there is an additional contribution from the lower film region, while the situation for 100~nm oxide depends on $\lambda_{ex}$.

From FIG.~\ref{interference} c,d) it can be seen that for $d=10$~nm (dash) and $d=70$~nm (dash-dot) $I(z)$ has a considerably high amplitude close to the substrate, extending some tens of nanometers into the film. As stated above, within this region a rearrangement of the 8CB molecules from random planar at the substrate~\cite{mullin_properties_1989} to smectic-A layers with the LC director aligned along the interface normal should take place~\cite{designolle_afm_2006}. We therefore expect the experimental FCS curves to contain some information about this disordered region and its extension $\zeta$. While for $d=10$~nm, there are two short regions (0~nm~$<z<40$~nm and $z>80$~nm) with high amplitude of $I(z)$, for $d=70$~nm, the amplitude is high in one large region (0~nm~$<z<80$~nm). In the latter case, vertical diffusion will cause slower fluorescence fluctuations in respect to the situation on $10$~nm oxide, leading to an enhanced FCS amplitude for slower correlation times for $d=70$~nm, which indeed can be seen in the experimental curves for correlation times $10^{-3}$~s~$<\tau<1$~s, compare FIG.~\ref{10nm} b) and~\ref{70nm} b).

\subsubsection{8CB films on 10 nm oxide - highlighting the disordered region}
\begin{figure}[ht]
  \includegraphics[width=2.5in]{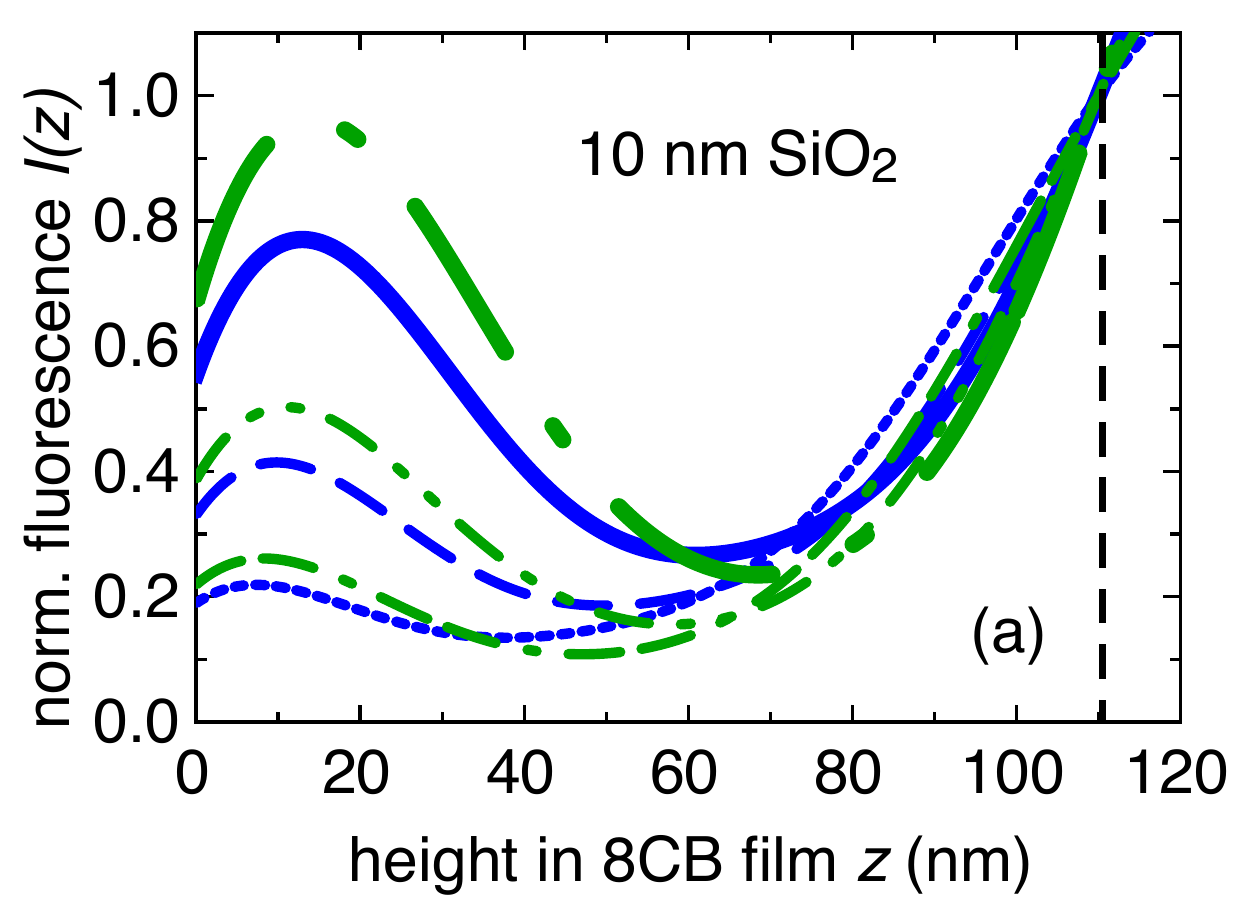}
 \includegraphics[width=2.5in]{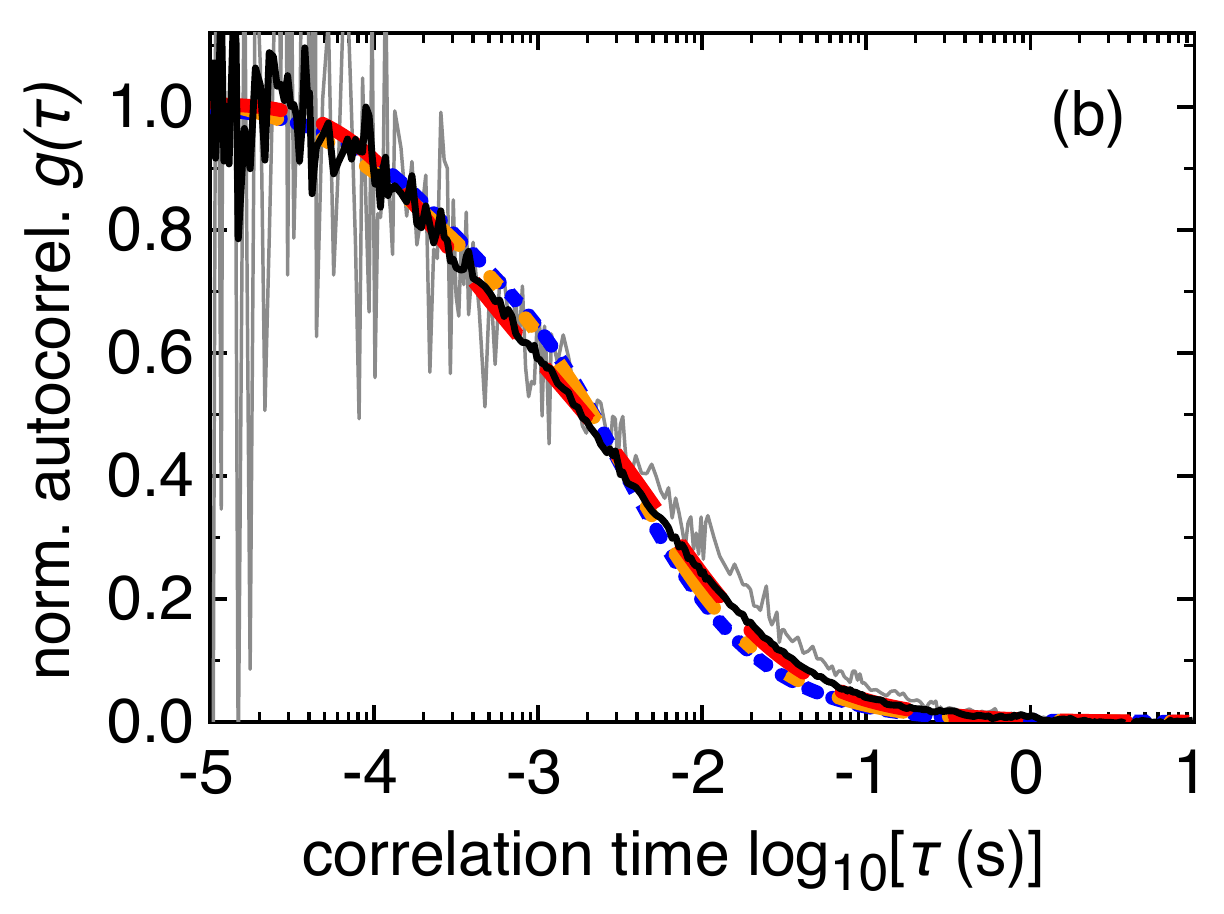}
  \includegraphics[width=1.8in]{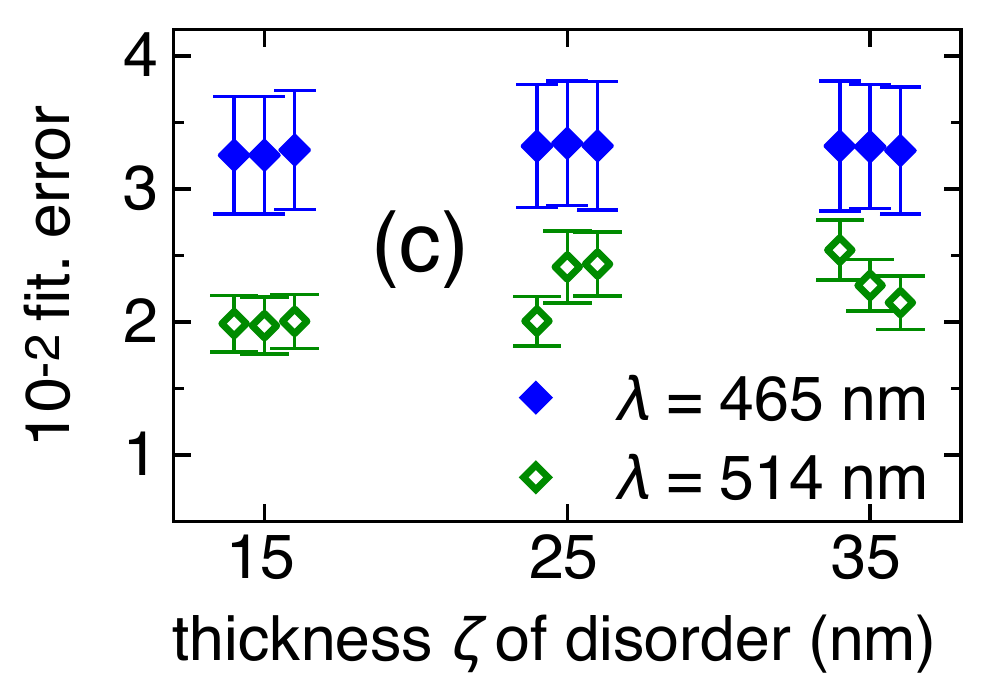}
  \includegraphics[width=3.2in]{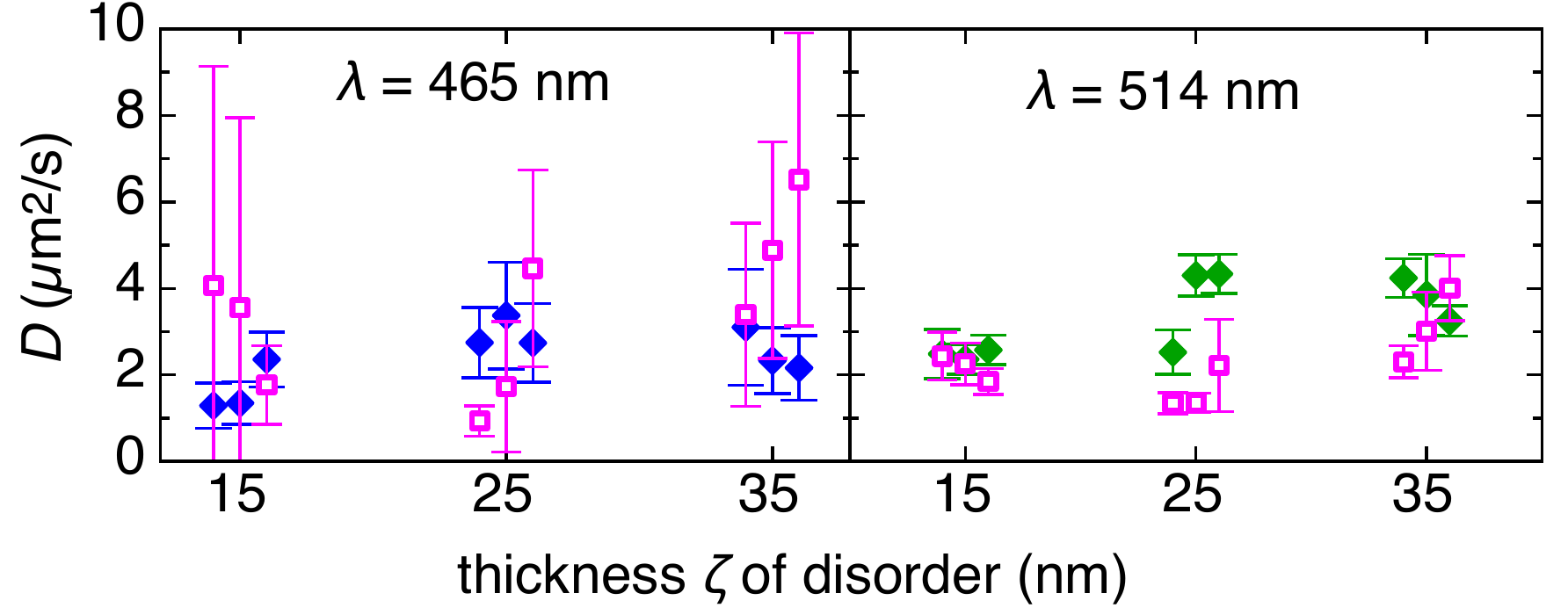}
\caption{\label{10nm} Results for 8CB films on 10~nm $\rm SiO_2$. a) $I(z)$ calculated for $\lambda_{ex}=465$~nm (blue) with $\zeta=$~15~nm (solid),  $\zeta=$~25~nm (dash), and $\zeta=$~35~nm (dot), as well as for $\lambda_{ex}=514$~nm (green) with $\zeta=$~15~nm (dash-dot), $\zeta=$~25~nm (dot-dash-dot), and $\zeta=$~35~nm (dash-dot-dash). b) Experimental FCS curves for $\lambda_{ex}=465$~nm (light solid) and $\lambda_{ex}=515$~nm (black solid) with fits for $\lambda_{ex}=515$~nm and $L=110$~nm and $\zeta=$~15~nm (red, dash), $\zeta=$~25~nm (blue, dot) and $\zeta=$~35~nm (yellow, dot-dash-dot). c) Fitting errors for $\lambda_{ex}=465$~nm ($\blacklozenge$) and $\lambda_{ex}=515$~nm ($\lozenge$). d,e) Diffusion coefficients  $D_{xy}$ ($\blacklozenge$) and $D_{z}$ ($\square$) obtained from fits. c-e) Values obtained for $L=100$~nm, $L=110$~nm, and $L=120$~nm displayed with 1~nm spacing. Error bars denote standard deviations using multiple data sets.}
\end{figure}
For both excitation wavelengths on 10~nm oxide, $I(z)$ is low in the middle of the films ($z\approx 60$~nm) with larger amplitudes for the lower and upper parts of the films, see FIG.~\ref{10nm}~a). The similar $I(z)$ result in similar FCS curves as can be seen in FIG.~\ref{10nm} b). The larger noise for $\lambda_{ex}=465$~nm (light solid) is due to the weaker absorption of PDI for that wavelength and results in a larger scatter of $D_z$, see FIG.~\ref{10nm}~d,~$\square$), because of the shorter correlation times related to vertical diffusion. For $\lambda_{ex}=515$~nm best fits were obtained for $L=110$~nm and $\zeta=$~15~nm (FIG.~\ref{10nm}~b, red dash). Fits with $\zeta=$~25~nm (blue dot) and $\zeta=$~35~nm (yellow dot-dash-dot) showed slightly larger deviations. This matches the situation for $\lambda_{ex}=465$~nm, if only the part of the FCS curve for $\tau>1$~ms is taken into account (fits not shown). 

For the upper film region ($z>80$~nm), the film consists of smectic~A layers, which leads to $D_{xy}=D_{\perp}<D_{\parallel}=D_z$. However, in the disordered region at the substrate, the diffusion anisotropy should lead to $D_{xy}\geq D_z$, which may even be enhanced by hydrodynamics showing a stronger decrease of the vertical diffusion coefficient, compare (\ref{Dpll_noslip}) to (\ref{Dperp_noslip}). The first peak of $I(z)$ contains the disordered region, but for $z>\zeta$ also smectic layers parallel to the interface, where $D_{xy}<D_z$. As can be seen in FIG.~\ref{10nm} a), for increasing $\zeta$ the amplitude $I(z)$ for the first peak ($z<40$)~nm decreases, while the amplitude for the upper film region ($z>80$)~nm stays almost the same. Therefore, the contribution from the disordered region to the FCS curve decreases with increasing $\zeta$, and we would expect $D_{xy}\approx D_z$ for $\zeta=$~15~nm, but  $D_{xy}<D_z$ for larger $\zeta$. For $\lambda_{ex}=515$~nm and for $\zeta=$~15~nm, $D_{xy}$ ($\blacklozenge$) and $D_z$ ($\square$) are approximately equal, matching the expectation, see FIG.~\ref{10nm} e). However, for $\zeta=$~25~nm, $D_{xy}>D_z$ contrary to the expectation, apart from $\zeta=$~35~nm, where $D_{z}$ is slightly larger than $D_{xy}$ for $L=120$~nm. But for this $\zeta$, the amplitude of $I(z)$ in the disordered region is very small and $D_{z}$ should be considerably larger than $D_{xy}$. Thus, the best fits and the obtained diffusion coefficients both point to the size of the disordered region to be in the range of 15~nm. The variation of film thickness $L$ within fitting had only minor effects in case of $\lambda_{ex}=515$~nm, while for $\lambda_{ex}=465$~nm most reasonable diffusion coefficients were obtained for $L=120$~nm pointing to a slightly thicker film for that particular sample.

\subsubsection{8CB films on 70 nm oxide - highlighting the disordered region}
\begin{figure}[htb]
  \includegraphics[width=2.5 in]{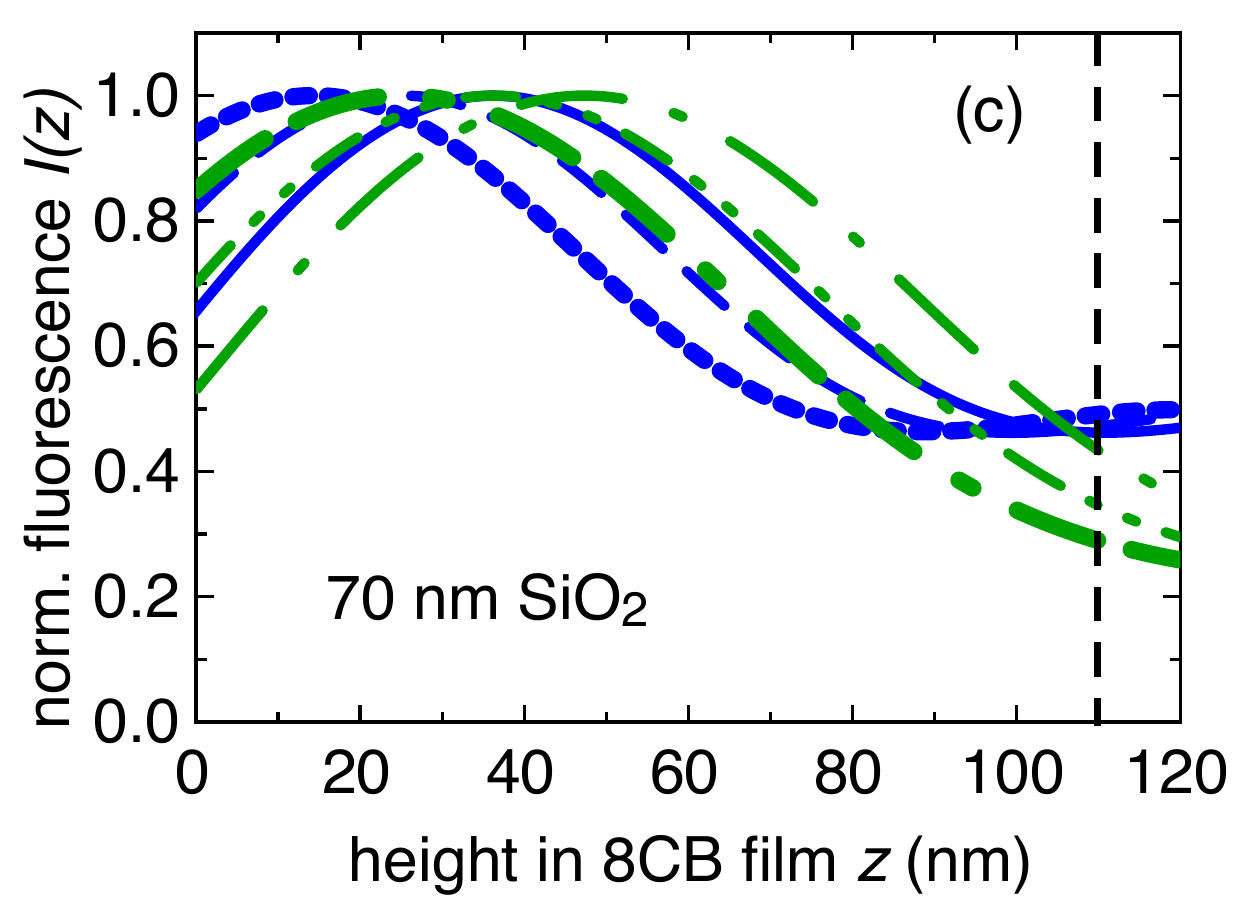}
  \includegraphics[width=2.5in]{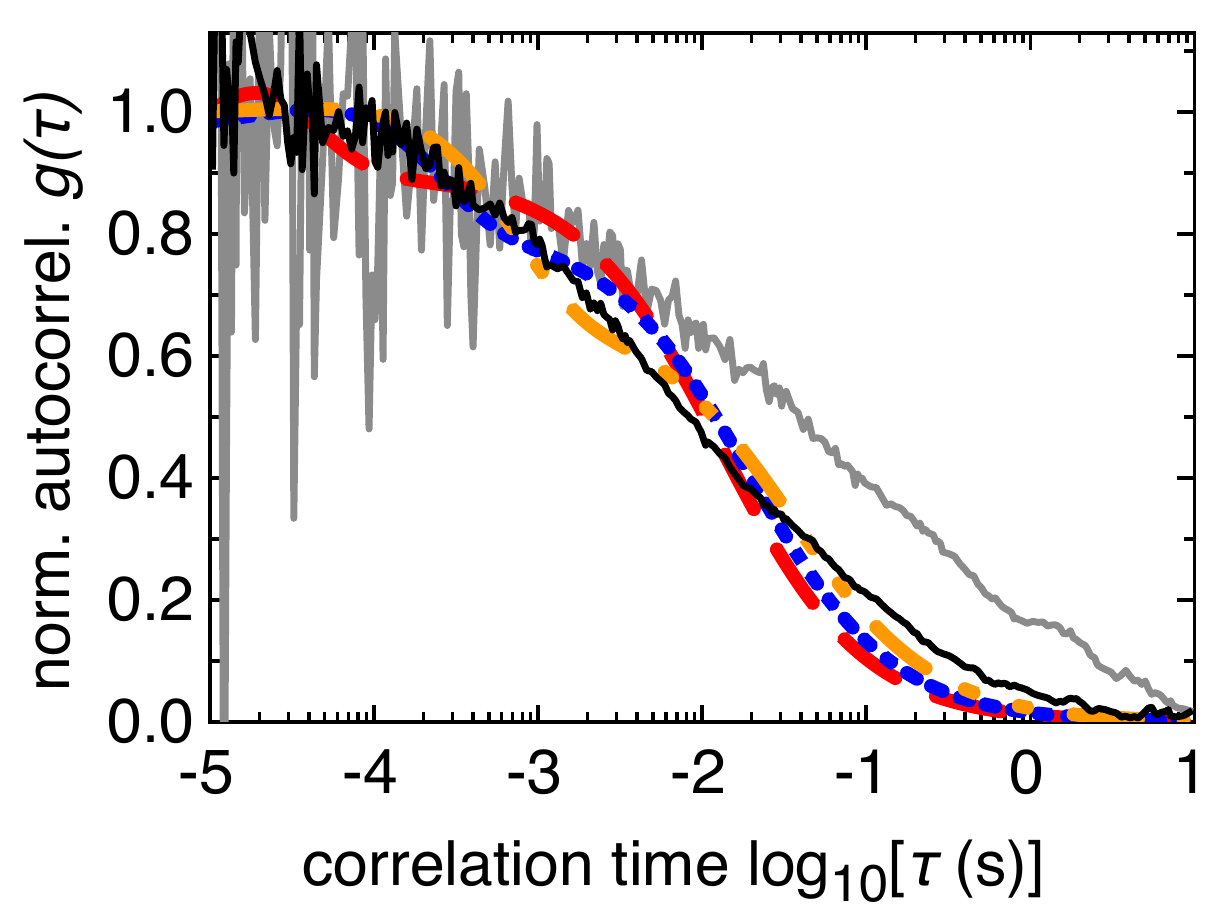}
 \includegraphics[width=1.8in]{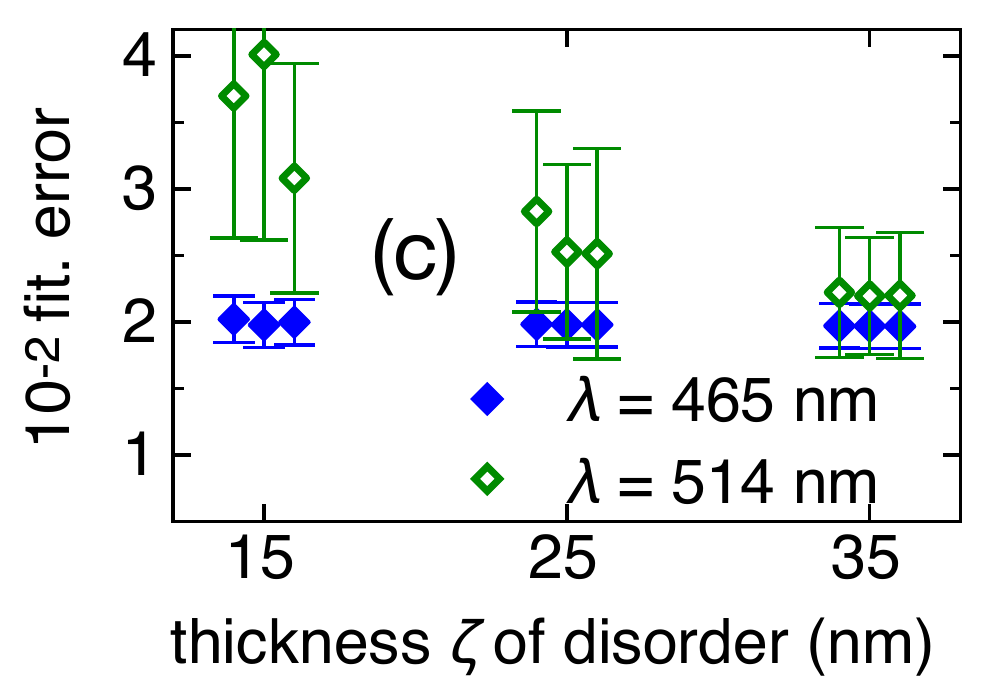}
   \includegraphics[width=3.2in]{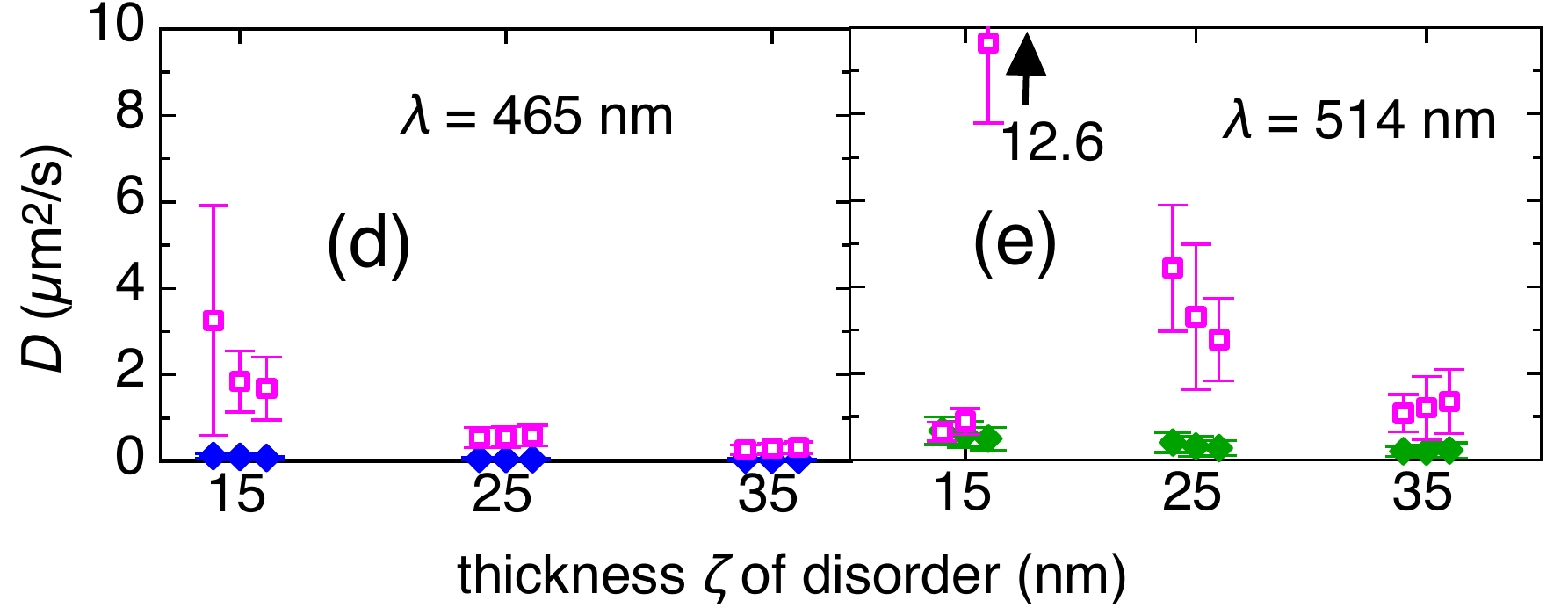}
\caption{\label{70nm} Results for 8CB films on 70~nm $\rm SiO_2$. a) $I(z)$ calculated for $\lambda_{ex}=465$~nm (blue) with $\zeta=$~15~nm (solid),  $\zeta=$~25~nm (dash), and $\zeta=$~35~nm (dot), as well as for $\lambda_{ex}=514$~nm (green) with $\zeta=$~15~nm (dash-dot), $\zeta=$~25~nm (dot-dash-dot), and $\zeta=$~35~nm (dash-dot-dash). b) experimental FCS curves (solids) with fits for $\lambda_{ex}=515$~nm. c) fitting errors for $\lambda_{ex}=465$~nm ($\blacklozenge$) and $\lambda_{ex}=515$~nm ($\lozenge$). d,e) diffusion coefficients  $D_{xy}$ ($\blacklozenge$) and $D_{z}$ ($\square$) obtained from fits. c-e)Values obtained for $L=100$~nm, $L=110$~nm, and $L=120$~nm displayed with 1~nm spacing. Error bars denote standard deviations using multiple data sets.}
\end{figure}
As can be seen in FIG.~\ref{70nm}~a), on 70~nm oxide, $I(z)>0.5$ for $z<80$~nm, with a shift of the peak to smaller $z$ for $\lambda_{ex}=465$~nm (blue) compared to $\lambda_{ex}=514$~nm (green). Consequently, the FCS curves obtained at  $\lambda_{ex}=465$~nm contain a larger contribution for longer correlation times $\tau$, related to slower diffusion close to the substrate, see FIG.~\ref{70nm}~b). Best fits were obtained for $\zeta=$~35~nm, as can be seen in FIG.~\ref{70nm}~b) showing an experimental FCS curve for $\lambda_{ex}=515$~nm (black solid) together with fits for $L=110$~nm and $\zeta=$~35~nm (yellow, dot-dash-dot),  $\zeta=$~25~nm (blue, dot), and $\zeta=$~15~nm (red, dash). $\zeta=$~35~nm also yielded the smallest errors with the most narrow standard deviations, see FIG.~\ref{70nm} c).

The high amplitude of $I(z)$ for $z<80$~nm covers the disordered region at the solid-liquid interface, as well as some part of the adjacent film region organized in smectic~A layers. For both excitation wavelengths, $I(z)$ calculated for $\zeta=$~35~nm yields the highest relative contribution from the disordered region, see FIG.~\ref{70nm}~a) (blue dot) for $\lambda_{ex}=465$~nm, and (green dash-dot-dash) for $\lambda_{ex}=514$~nm. As discussed previously, in the disordered region we expect $D_z\leq D_{xy}$ and in the smectic~A layers $D_z> D_{xy}$, which here should lead to $D_z\geq D_{xy}$, due to the  larger contribution from the smectic~A region. Fitting results match this expectation, see (FIG.~\ref{70nm}~d,e), with only minor influence from variation of $L$. For $\lambda_{ex}=514$~nm, most reasonable diffusion coefficients are obtained for $\zeta=$~35~nm, agreeing with best fits seen in FIG.~\ref{70nm}~b) (yellow, dot-dash-dot). For $\lambda_{ex}=465$~nm only 7~FCS curves could be analyzed, the results follow the trend seen for $\lambda_{ex}=514$~nm.

\subsubsection{8CB films on 25 nm oxide - upper film region only}
\begin{figure}[htb]
  \includegraphics[width=2.5in]{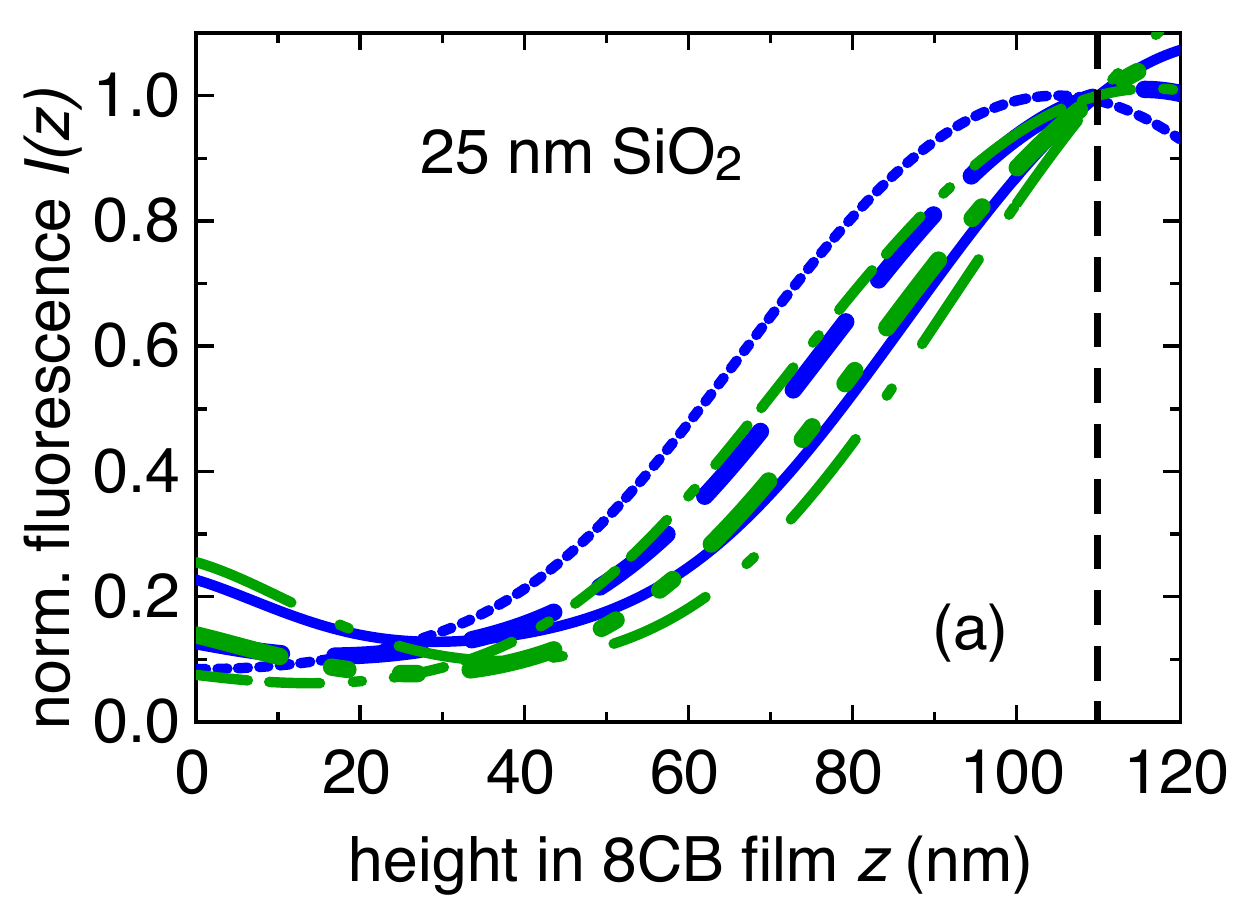}
 \includegraphics[width=2.5in]{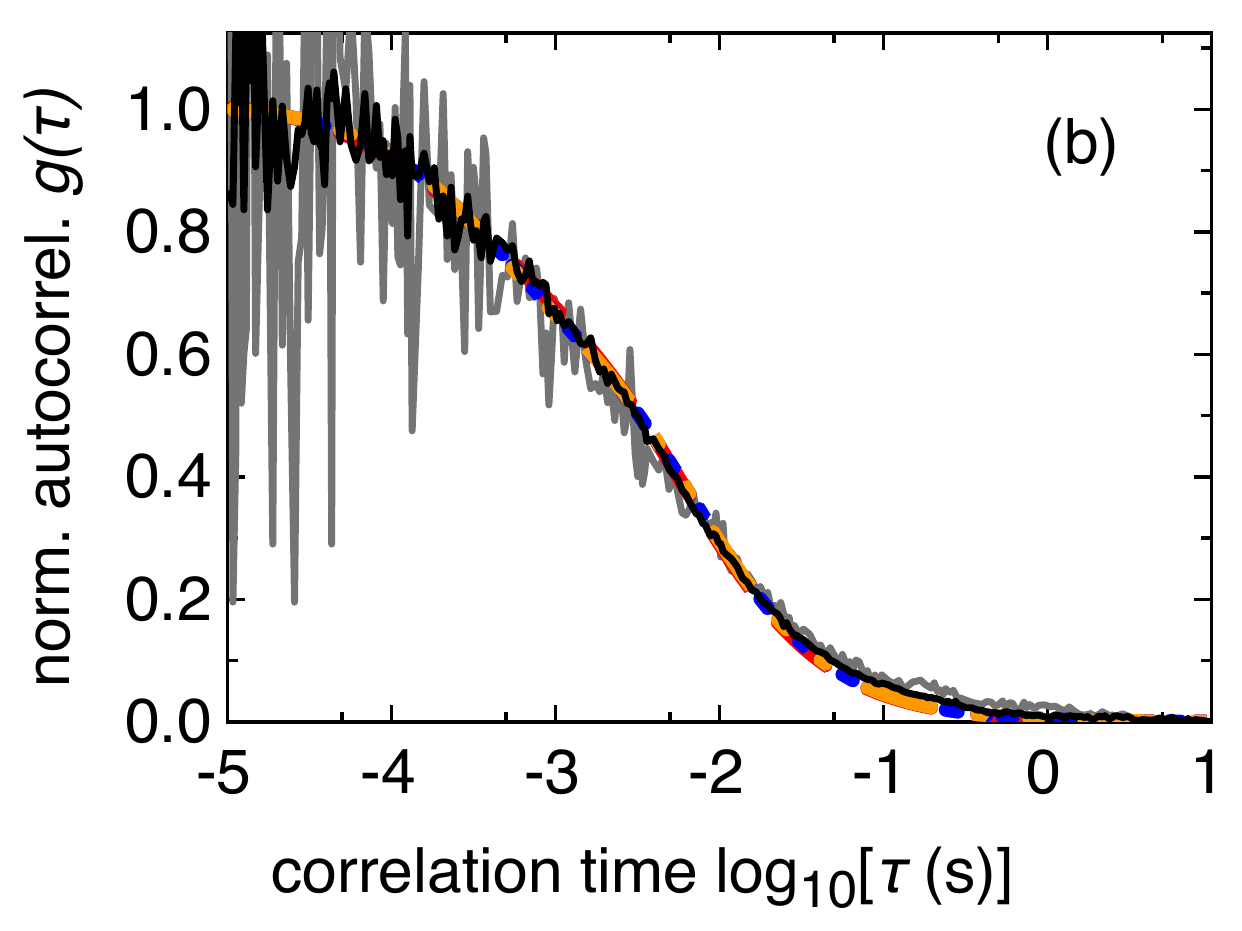}
  \includegraphics[width=1.8in]{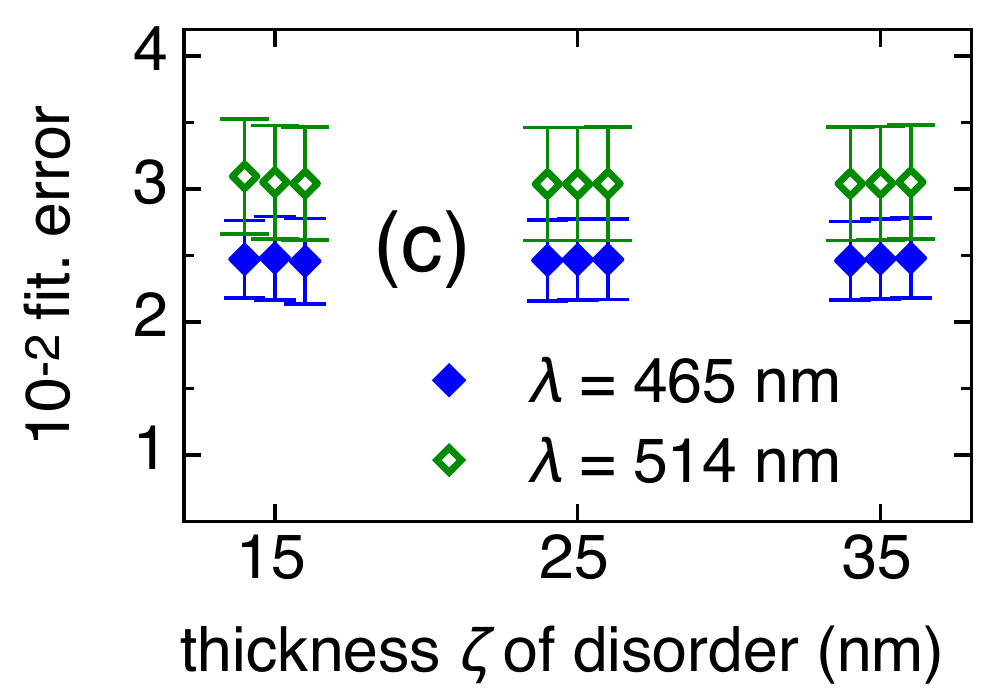}
  \includegraphics[width=3.2in]{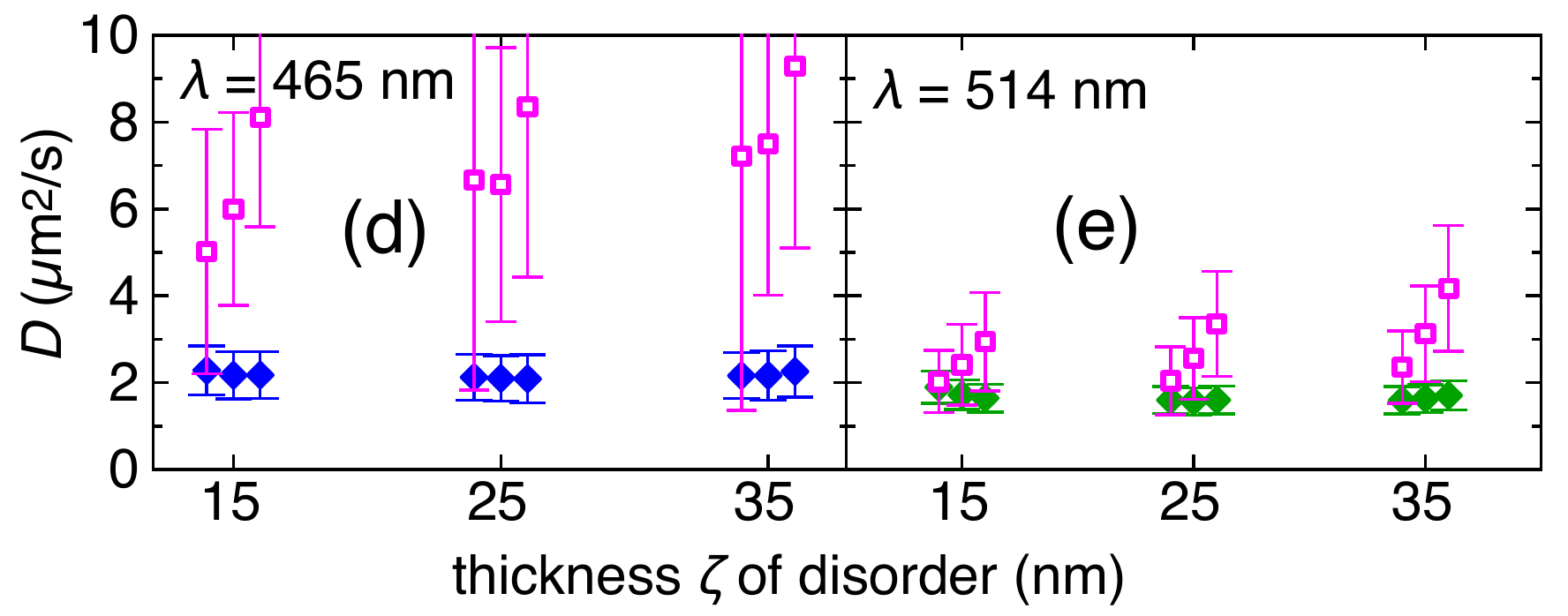}
\caption{\label{25nm} Results for 8CB films on 25~nm $\rm SiO_2$. a) $I(z)$  calculated for $\lambda_{ex}=465$~nm (blue) with $\zeta=$~15~nm (solid),  $\zeta=$~25~nm (dash), and $\zeta=$~35~nm (dot), as well as for $\lambda_{ex}=514$~nm (green) with $\zeta=$~15~nm (dash-dot), $\zeta=$~25~nm (dot-dash-dot), and $\zeta=$~35~nm (dash-dot-dash). b) experimental FCS curves (solids) with fits for $\lambda_{ex}=515$~nm. c) Fitting errors for $\lambda_{ex}=465$~nm ($\blacklozenge$) and $\lambda_{ex}=515$~nm ($\lozenge$). d,e) Diffusion coefficients  $D_{xy}$ ($\blacklozenge$) and $D_{z}$ ($\square$) obtained from fits. c-e) Values obtained for $L=100$~nm, $L=110$~nm, and $L=120$~nm displayed with 1~nm spacing. Error bars denote standard deviations using multiple data sets.}
\end{figure}
For $d=25$~nm all calculated $I(z)$ show a significant amplitude only for $z>70$~nm, see FIG.~\ref{25nm} a). Consequently, the fluorescence can be detected only from the upper part of the 8CB film leading to similar experimental FCS curves for both excitation wavelengths, see FIG.~\ref{25nm} b), and the fit results do not depend on $\zeta$, see FIG.~\ref{25nm} c-e). The rather large fitting errors are induced by the narrow vertical film region contributing to the signal. On short vertical distances, higher orders of the correlation function might have to be taken into account for improving the fits. The only small observed vertical area also leads to large standard deviations for the $D_i$, see FIG.~\ref{25nm}~d,e), which are most pronounced for $\lambda_{ex}=465$~nm and $D_{z}$, due to the large noise at short $\tau$ in that case (FIG.~\ref{25nm}~b)). Due to the only small upper vertical region contributing to the signal, the actual film thickness $L$ has a considerable influence on the obtained $D(z)$, as can be seen in FIG~\ref{25nm}~d,e).

\subsubsection{8CB films on native oxide - lower and upper film regions}
\begin{figure}[htb]
 \includegraphics[width=2.5in]{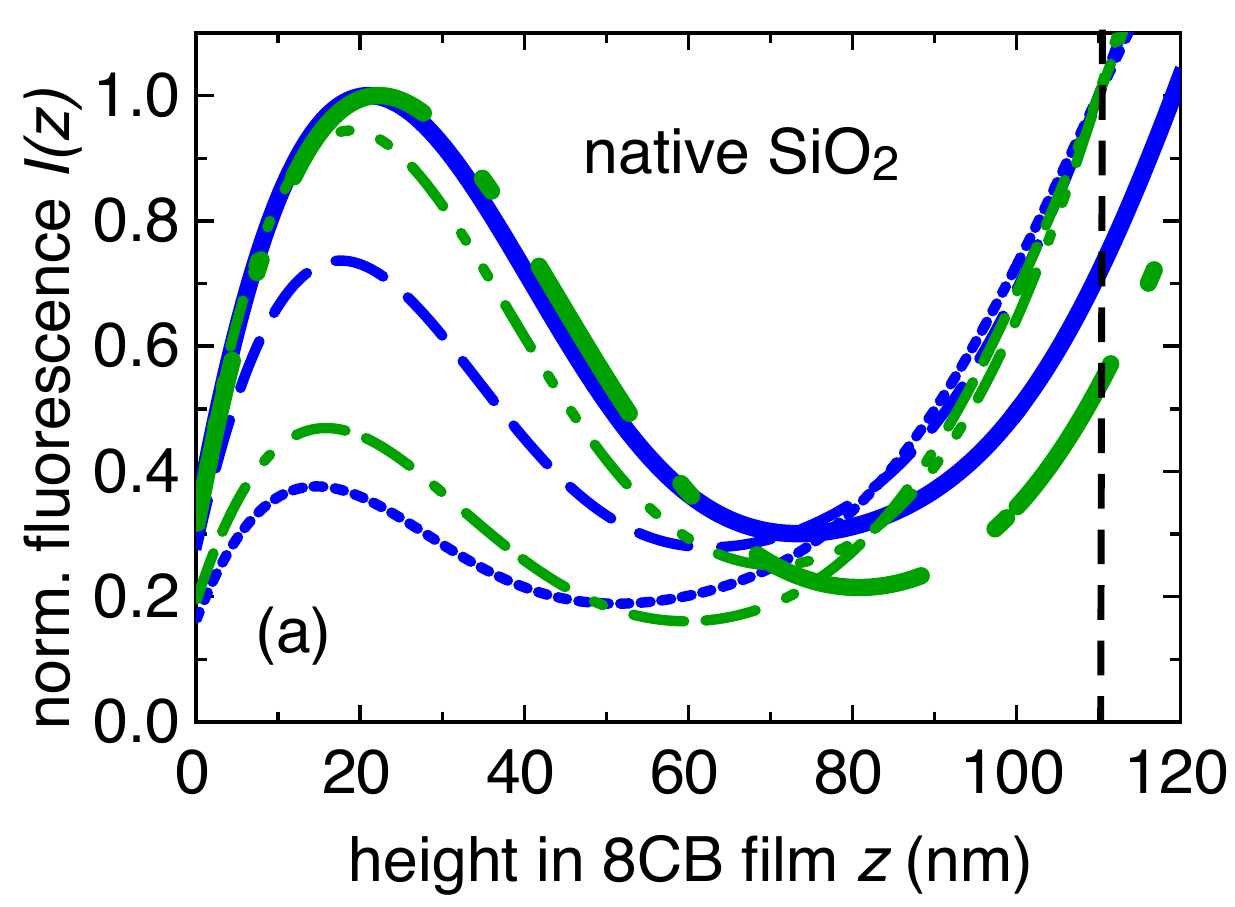}
 \includegraphics[width=2.5in]{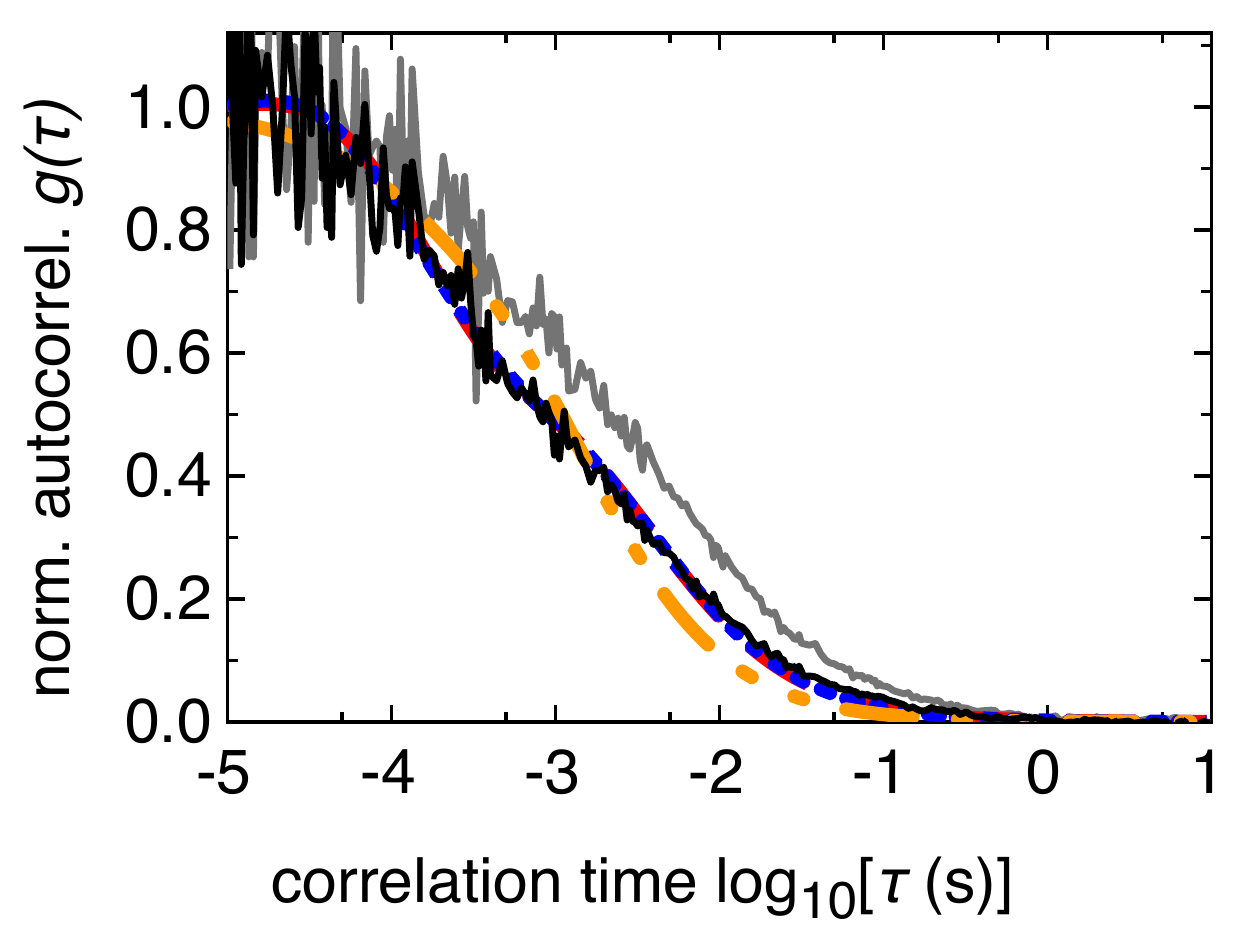}
 \includegraphics[width=1.8in]{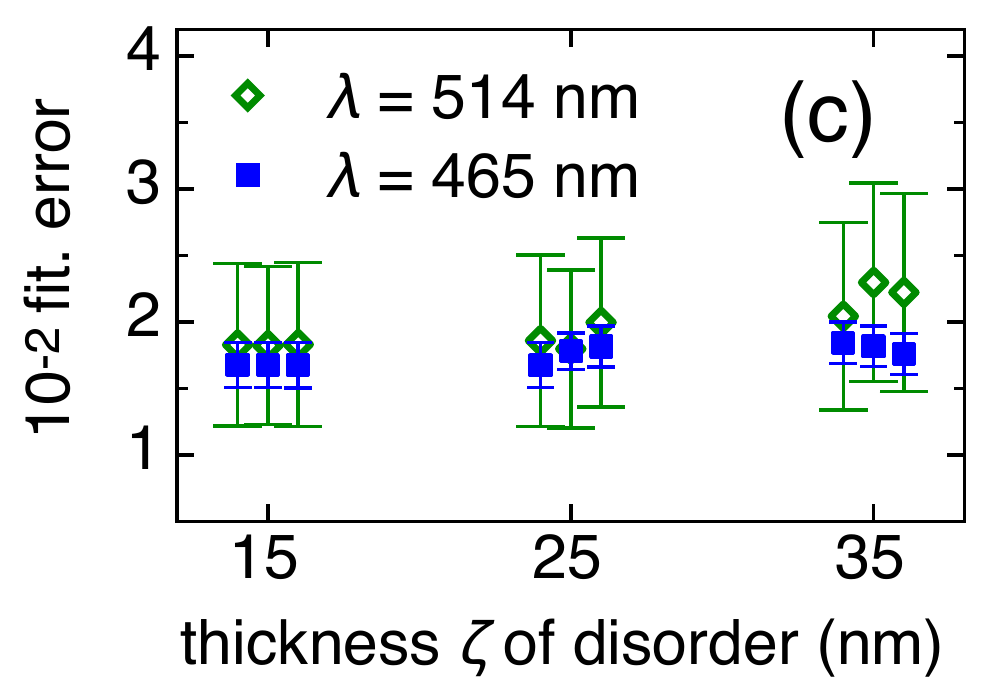}
 \includegraphics[width=3.2in]{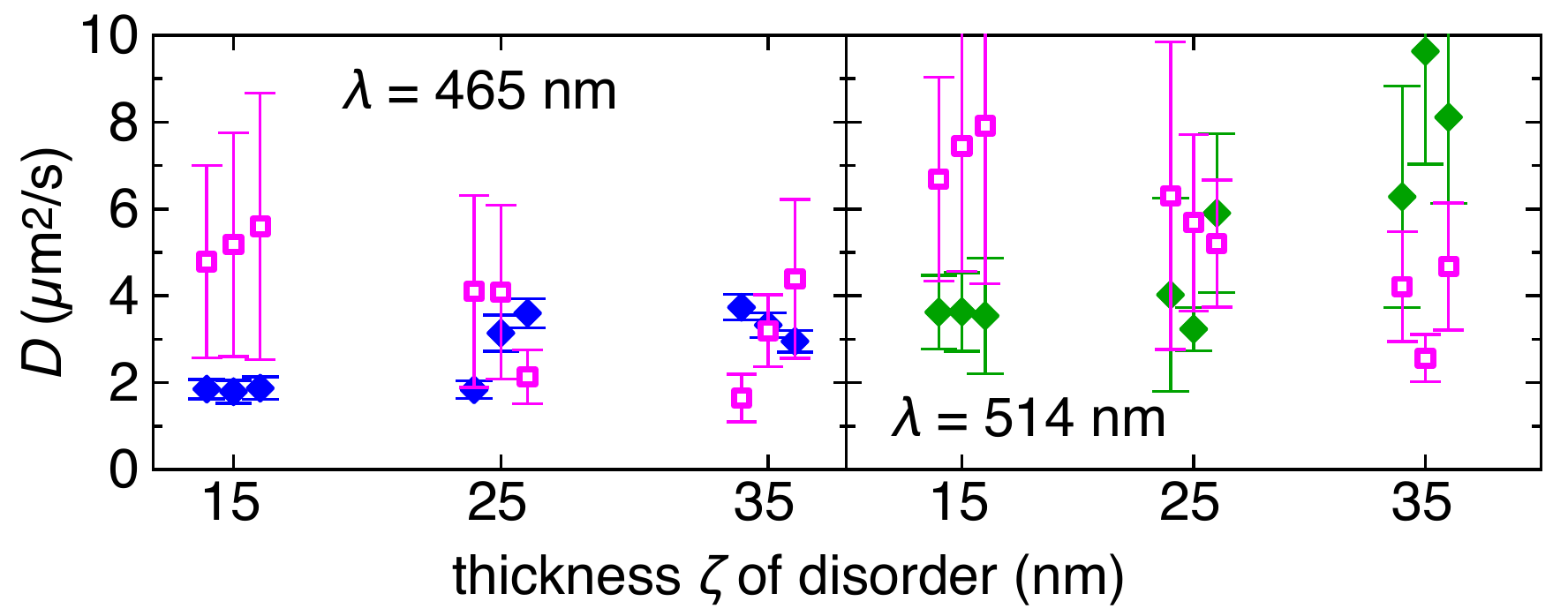}
\caption{\label{4nm} Results for 8CB films on native $\rm SiO_2$. a) $I(z)$ calculated for $\lambda_{ex}=465$~nm (blue) with $\zeta=$~15~nm (solid),  $\zeta=$~25~nm (dash), and $\zeta=$~35~nm (dot), as well as for $\lambda_{ex}=514$~nm (green) with $\zeta=$~15~nm (dash-dot), $\zeta=$~25~nm (dot-dash-dot), and $\zeta=$~35~nm (dash-dot-dash). b) experimental FCS curves (solids) with fits for $\lambda_{ex}=515$~nm. c) fitting errors for $\lambda_{ex}=465$~nm ($\blacklozenge$) and $\lambda_{ex}=515$~nm ($\lozenge$). d,e) diffusion coefficients  $D_{xy}$ ($\blacklozenge$) and $D_{z}$ ($\square$) obtained from fits. c-e) Values obtained for $L=100$~nm, $L=110$~nm, and $L=120$~nm shown with 1~nm spacing. Error bars give standard deviations using multiple data sets.}
\end{figure}
Similar to the situation for $d=10$~nm, on native oxide $I(z)$ contains two peaks for $z\leq L$, but there is a considerably smaller contribution to $I(z)$ from the close proximity to the solid-liquid interface ($z<5$~nm), see FIG~\ref{4nm}~a) and thus from the disordered region. We therefore expect the diffusion coefficients to be larger than for $d=10$~nm, with $D_z>D_{xy}$. Also here best fits are obtained for $\zeta=15$~nm, as can be seen exemplarily in FIG~\ref{4nm}~b). Fitting errors are similar within standard deviations from multiple data sets (FIG.~\ref{4nm}~c). For $\lambda_{ex}=514$~nm and $\zeta=35$~nm we obtained $D_{xy}\gg D_{z}$, which is unreasonable. For $\zeta=25$~nm, $D_{xy}\approx D_{z}$ within standard deviation, while only for $\zeta=15$~nm, $D_{xy}< D_{z}$, as expected. Thus, also on native oxide, best fits and the obtained diffusion coefficients both point to $\zeta=15$~nm. The larger standard deviations obtained for $D_{z}$, see FIG.~\ref{4nm}~d,e), are caused by the noisy FCS curves at short $\tau$ for both $\lambda_{ex}$, see FIG.~\ref{4nm}~b), and the narrow $z\gtrsim100$~nm where $I(z)>0.5$.

\subsubsection{8CB films on 100 nm oxide - highlighting the solid-liquid interface}
\begin{figure}[htb]
 \includegraphics[width=2.5in]{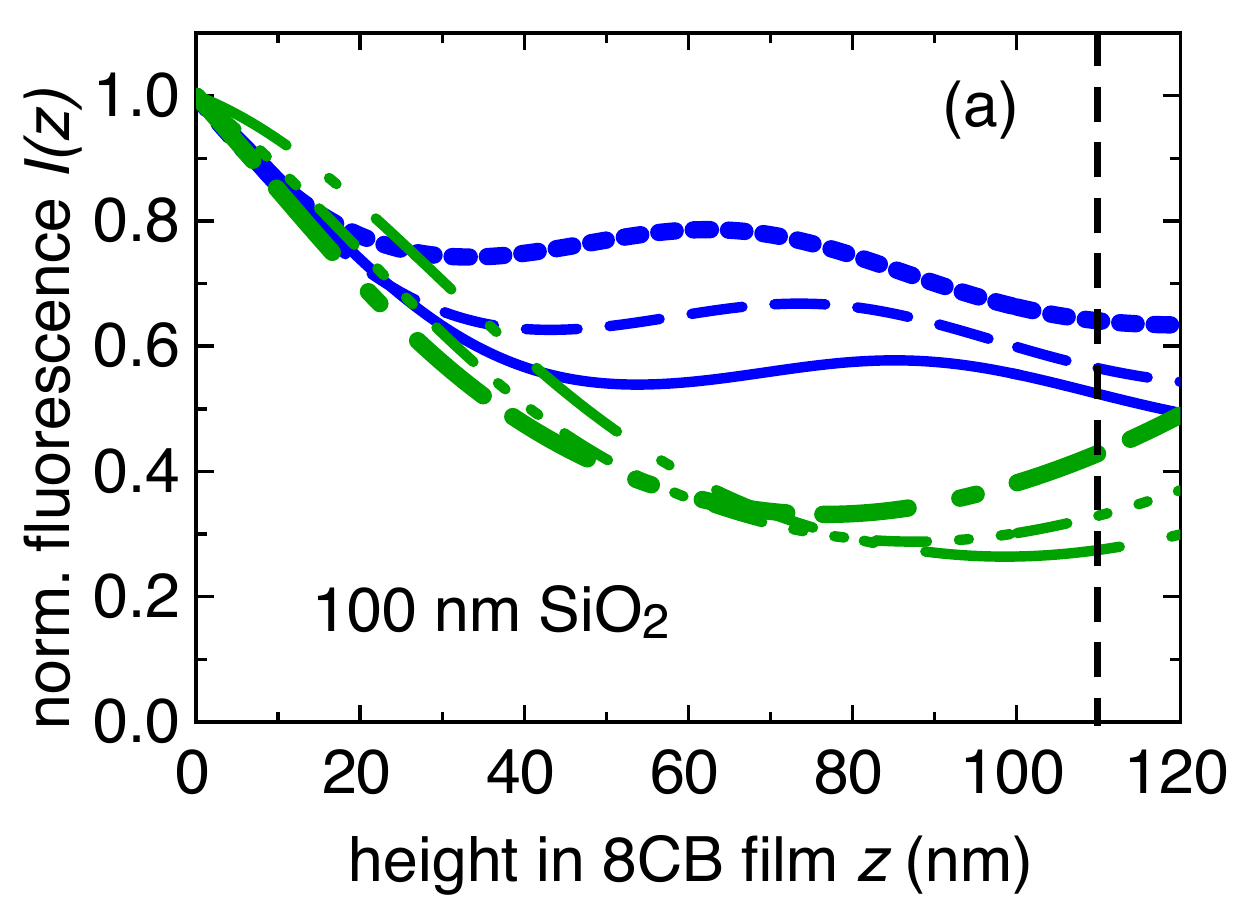}
 \includegraphics[width=2.5in]{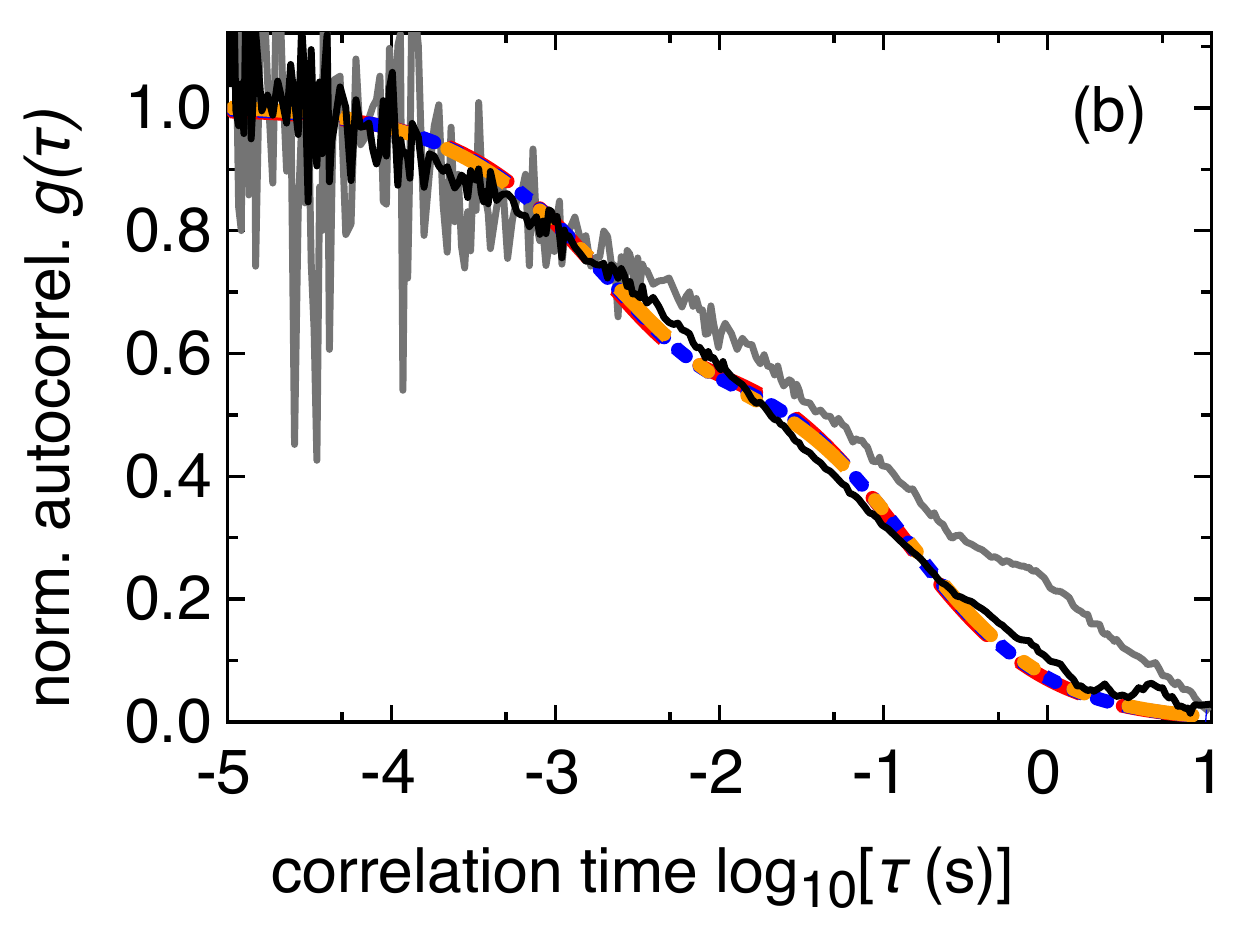}
 \includegraphics[width=1.7in]{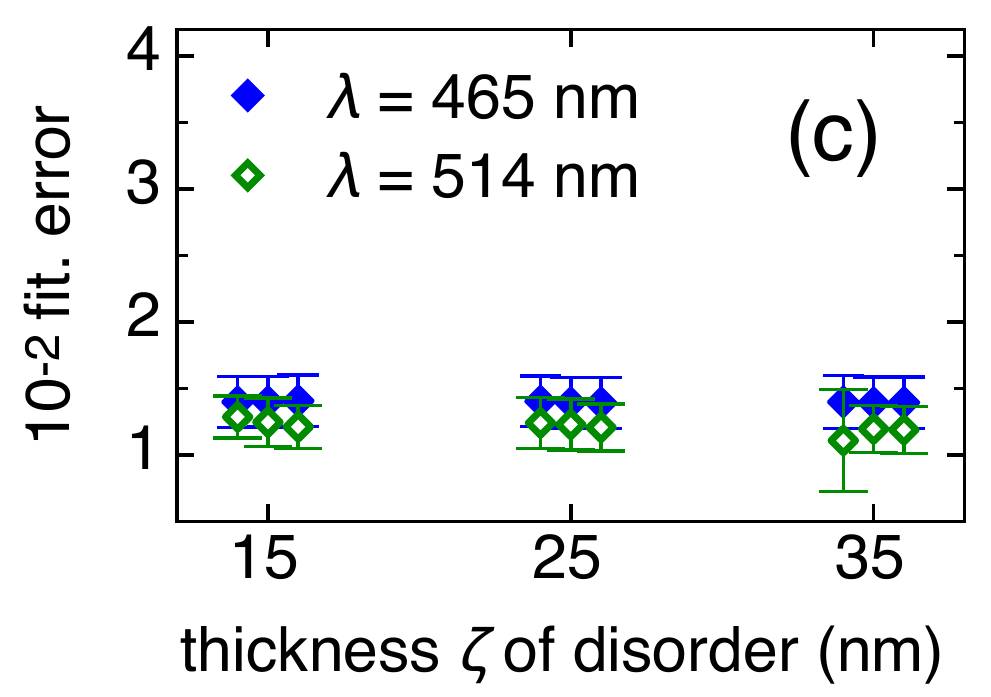}
 \includegraphics[width=3.1in]{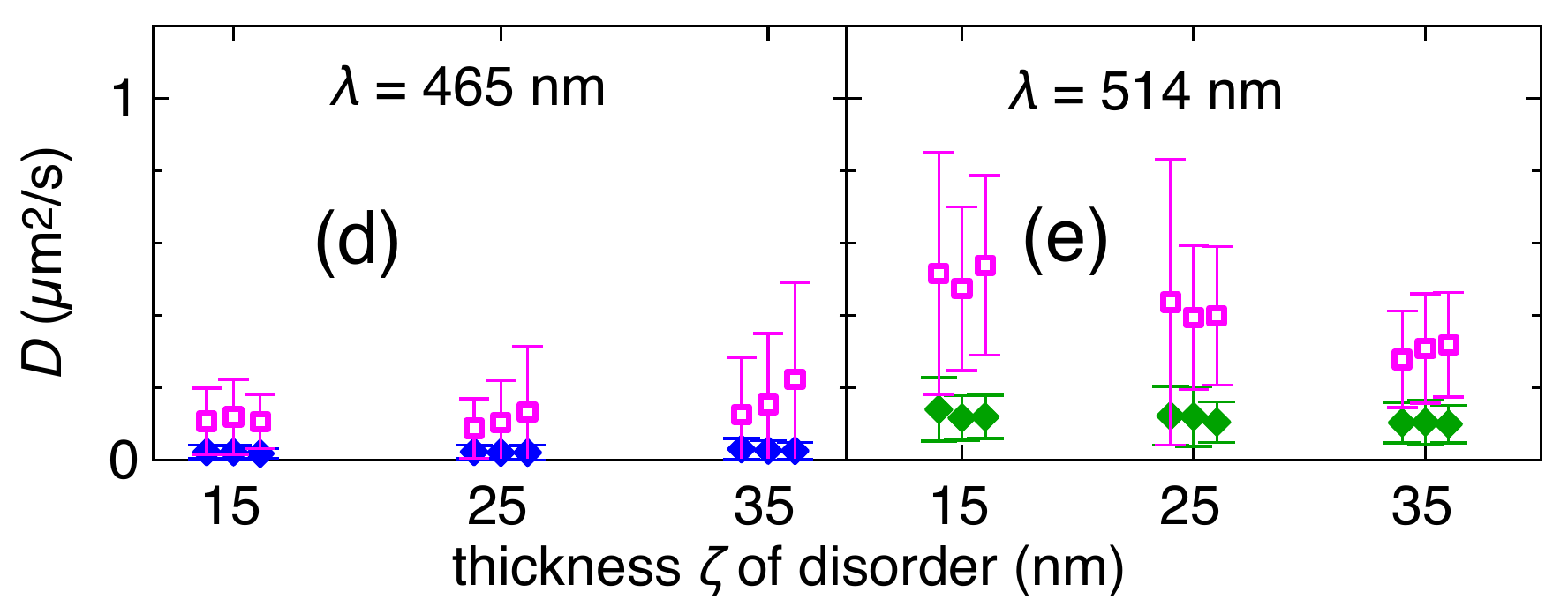}
\caption{\label{100nm} Results for 8CB films on 100~nm $\rm SiO_2$. a) $I(z)$  calculated for $\lambda_{ex}=465$~nm (blue) with $\zeta=$~15~nm (solid),  $\zeta=$~25~nm (dash), and $\zeta=$~35~nm (dot), as well as for $\lambda_{ex}=514$~nm (green) with $\zeta=$~15~nm (dash-dot), $\zeta=$~25~nm (dot-dash-dot), and $\zeta=$~35~nm (dash-dot-dash). b) experimental FCS curves (solids) with fits for $\lambda_{ex}=515$~nm. c) fitting errors for $\lambda_{ex}=465$~nm ($\blacklozenge$) and $\lambda_{ex}=515$~nm ($\lozenge$). d,e) diffusion coefficients  $D_{xy}$ ($\blacklozenge$) and $D_{z}$ ($\square$) obtained from fits, note the smaller scale. c-e) Values obtained for $L=100$~nm, $L=110$~nm, and $L=120$~nm displayed with 1~nm spacing. Error bars denote standard deviations using multiple data sets.}
\end{figure}
As can be seen in FIG.~\ref{100nm}~a), for $d=100$~nm the general shape of $I(z)$ obtained for different $\zeta$ is similar for each excitation wavelength, which leads to similar fitting errors (FIG.~\ref{100nm}~c)) and only minor variations in diffusion coefficients (FIG.~\ref{100nm}~d,e), zoomed in scale!), rendering a discrimination of $\zeta$ from this experiment not feasible. 

A comparison of calculated $I(z)$, see FIG.~\ref{interference} c,d), yields the highest influence from the solid-liquid interface for $d=100$~nm. This is most prominent for $\lambda_{ex}=514$~nm, because here $I(z)$ decreases to about $40\%$ of its initial value within $40$~nm distance to the substrate (FIG.~\ref{100nm}~a, dash-dot-dash). In contrast, for $\lambda_{ex}=465$~nm and in particular for $\zeta=35$~nm, $I(z)>0.6$ in the entire film (FIG.~\ref{100nm}~a, blue dot). This leads to an only small intensity of fluorescence fluctuations for vertical diffusion, which in turn causes a small absolute amplitude of the FCS curves, see FIG~\ref{interference}~b,~$\blacktriangle$), approaching the situation on glass substrates (FIG~\ref{interference}~b,~o). The smaller influence from vertical diffusion also leads to an increase in the relative amplitude for longer $\tau$ in the normalized FCS curve obtained at $\lambda_{ex}=465$~nm, compared to the curve at $\lambda_{ex}=514$~nm, see FIG.~\ref{100nm}~b) light solid and black solid, respectively. For $\lambda_{ex}=465$~nm, fitting renders $D_{xy}$ and $D_{z}$ of similar size as their standard deviations from multiple fitting, therefore, a further discussion is not feasible.

\begin{acknowledgments}
This work was funded by the German Research Foundation (DFG) within FOR 877 "From local constraints to macroscopic transport" and a personal research grant Ta 1049/1-1 to D.T., and performed in the context of the European COST Action MP1302 Nanospectroscopy. 
\end{acknowledgments}

\bibliography{FCSreflection}

\end{document}